\documentclass[journal]{IEEEtran}

\usepackage{cite}

   \usepackage[pdftex]{graphicx}
   \usepackage{epsfig}
   \DeclareGraphicsExtensions{.eps,.png}
\usepackage[cmex10]{amsmath}
\usepackage{amssymb}

\usepackage{array}

\usepackage[caption=false,font=footnotesize]{subfig}

\usepackage{dblfloatfix}

\usepackage{url}

\usepackage{pgf,tikz,tikz-3dplot,pgfplots} 
\usepackage{tkz-euclide}
\usetikzlibrary{calc,decorations.markings,arrows,automata,backgrounds}
\usetikzlibrary{shapes.geometric,shapes,intersections,decorations.pathreplacing}
\usetkzobj{all}
\pgfplotsset{compat=newest}

\usepackage{braidtikz}

\usepackage[normalem]{ulem}
\usepackage{cancel}

\newtheorem{theorem}{Theorem}

\newtheorem{definition}{Definition}[section]

\newtheorem{lemma}{Lemma}
\newtheorem{restr}{Restriction}

\newcommand{\R}{\mathbb{R}}
\newcommand{\T}{\mathbf{T}}
\newcommand{\argmin}{\operatorname*{arg\,min}}

\newcommand{\norm}[1]{\left\| #1 \right\|}
\newcommand{\abs}[1]{\left| #1 \right|}
\def\qed{\hfill\rule[-1pt]{5pt}{5pt}\par\medskip}

\begin{document}

\title{Inter-Robot Interactions in\\ Multi-Robot Systems Using Braids}

\author{Yancy~Diaz-Mercado~%
        and~Magnus~Egerstedt% <-this % stops a space
\thanks{The authors are is with the Department
of Electrical and Computer Engineering, Georgia Institute of Technology, Atlanta,
GA, 30332 USA. \mbox{E-mail: \{\texttt{yancy.diaz@gatech.edu},~\texttt{magnus@gatech.edu}\}.}}% <-this % stops a space
}

\maketitle

\begin{abstract}
This paper describes a framework for multi-robot coordination and motion planning with emphasis on inter-agent interactions. We focus on the characterization of inter-agent interactions with sufficient level of abstraction so as to allow for the enforcement of desired interaction patterns in a provably safe (i.e., collision-free) manner, e.g., for achieving rich movement patterns in a shared space, or to exchange sensor information. We propose to specify interaction patterns through elements of the so-called braid group. This allows us to not focus on a particular pattern per se, but rather on the problem of being able to execute a whole class of patterns. The result from such a construction is a hybrid system driven by symbolic inputs that must be mapped onto actual paths that both realize the desired interaction levels and remain safe in the sense that collisions are avoided.
\end{abstract}

%\begin{IEEEkeywords}
%Multi-robot motion planning, inter-robot interactions, hybrid controllers, braids, mixing, collision-free.
%\end{IEEEkeywords}

\section{Introduction}
\IEEEPARstart{M}{any} applications have been proposed for multi-robot systems. For example, the multi-robot foraging paradigm, in which agents wander around an environment searching for items of interest \cite{balch_foraging,
sugawara_foraging}, can relate to many real-world problems (e.g., waste or specimen collection in hazardous environments, explosive detection and disposal, and search and rescue operations after a crisis such as floods and earthquakes). 
Another application of interest in the research community and in industry is cooperative assembly \cite{bonert_assembly,
nguyen_assembly} and self-assembly \cite{gross_self_assembly,klavins_self_assembly}.
In the cooperative assembly scenario, multiple robots need to coordinate their motion in order to cooperatively assemble a structure using a possibly heterogeneous set of tools, end effectors, skills, or parts. Similarly, in the self-assembly scenario, multiple robots need to coordinate their motion in order to collectively form a structure or achieve a geometric configuration. 
There has been a recent push for robotic farming and precision agriculture 
\cite{pedersen_farming,noguchi_farming}, where a fleet of robots is sent to gather data on the status of crops, tend to and harvest them. 
In some communication architectures, mobile agents called data MULEs or message ferries \cite{DataMules, 
vasilescu_data_mules,zhao_data_mules} are used to transport data between sensors, access points, or base stations, in situations where it is impractical to establish physical communication. Multi-robot simultaneous localization and mapping (SLAM) takes advantage of the robot team's size to attain a more complete (or more detailed) map faster than with a single robot by combining information from multiple sources and coordinating motion \cite{howard_slam,
ozkucur_slam}. 
Other applications include transportation systems (e.g., intelligent highways, air traffic control) \cite{murray_transportation}, and the convoy protection scenario (e.g., surveillance, coverage) \cite{ding_convoy_protection}.

In many of these applications,
the overall objectives can be stated in terms of making a team of robots follow a physical path, such as a road or the movements of a ground convoy, while ensuring that particular search patterns are executed \cite{spiral1,line1,spiral2}. 
These patterns should be selected in such a way that certain secondary geometric objectives are met, including ensuring that an area along the path is covered, that multiple views of the same objects are achieved, that an aerial vehicle is always on top of the convoy, or that sufficiently many vehicle-to-vehicle interactions take place for the purpose of information sharing
\cite{KimNet09,Tovar}. 
In this work, we collect all of these different secondary objectives under one unified banner, namely \emph{multi-robot mixing}. In particular, we specify interaction patterns with certain desired levels of mixing, and then proceed to generate the actual cooperative movements that realize these mixing levels.

In this paper we study the problem of characterizing inter-robot interactions for the sake of coordination and collision avoidance. We specify the mixing patterns through elements of the so-called braid group
\cite{braid1,braid2}, 
where each element corresponds to a particular mixing pattern, i.e., we do not focus on a particular pattern {\it per se}, but rather on the problem of being able to execute a whole class of patterns. The result from such a construction is a hybrid system driven by symbolic inputs 
\cite{SymbolicPlanning}, 
i.e., the braids, that must be mapped onto actual paths that both obtain the mixing level specified through the braid, and remain safe in the sense that collisions are avoided.

The use of symbolic inputs allows us to abstract away the geometry and physical constraints involved in the multi-robot motion planning. As described in \cite{SymbolicPlanning}, this provides the advantage of hierarchical abstractions
that are
typically broken into three stages. At the top layer is the \emph{specification level}, which describes the motion tasks (such as robots \textit{A} and \textit{B} should interact, and arrive at the goal simultaneously). The second layer is the \emph{execution level}, which describes \emph{how} to obtain the motion plans, e.g., by generating trajectories based on optimality conditions. The bottom layer is the \emph{implementation level}, which concerns itself with constructing the robot controller, e.g., to track a reference trajectory.
This work also extends the notions presented in \cite{egerstedt_virtual_vehicles} for trajectory tracking of virtual vehicles. The idea there is to have the physical vehicles track a virtual vehicle, as opposed to a path (or trajectory) itself. The virtual vehicle, which is being controlled directly with simplified dynamics, can be controlled in order to track a reference path while satisfying the constraints, compensating for the physical vehicle's dynamics and other disturbances.

The outline of the paper is as follows: in Section 
\ref{sec:braid}
we start with
a brief summary of the existing literature on braiding for the sake of multi-robot motion planning;
in Section \ref{sec:GeomInterp}, the braid group is introduced as a way of specifying mixing levels and the corresponding symbolic braid objects are given a geometric interpretation in terms of planar robot paths; controllers are then proposed so that mixing strategies satisfying a given specification can be executed by a class of robots as described in Section \ref{sec:braid_controllers}, together with a bound on the highest achievable mixing level.  
In Section \ref{sec:implementing_braids}, implementation of these controllers on actual robotic platforms is addressed, 
and these ideas are deployed on a team of actual mobile robots.

%%%%%%%%%%%%%%%%%%%%%%%%%%%%%%%%%%%%%%%%%%%%%%%%%%%%%%%%%%%%%%%%%%%%%%%%%
\section{Braids and Robotics}
\label{sec:braid}

The use of braids for robot motion planning has been considered before. Using the notion of configuration space \cite{configuration_space} for robot motion planning, \cite{ghrist_braids} studies the problem of construction and classification of configuration spaces for graphs, e.g., robots on a manufacturing floor constrained on rails or paths. By studying the topological data associated with these graphs, such as the braid groups, he is able to provide a measure of the complexity of the control problem (e.g., the construction of potential field controllers on homeomorphic spaces).

In \cite{Kurlin2009}, the \emph{graph braid group} is used as the fundamental group of the configuration space of graphs that describe robot motion. There, each graph in the configuration space represents discretized collision-free robotic motion plan (e.g. road maps), where at each discrete time the graph vertices represent the positions of robots and (possibly moving) obstacles, and the edges represent fixed tracks connecting these vertices. 
They present an algorithm to construct the presentation of the graph braid group of $N$ agents, where the group generators (i.e., the braids) represent actual paths between configurations of robots on the graph, i.e., the motion plan to transition from one configuration to another. 
However, they only consider zero-size robots where they rely on a ``one edge separation'' between points at all times to avoid collisions, and as such, their approach is purely academic and mainly focuses on the combinatorics of ideal robots moving on a graph.

We will use the braid group's generators as symbolic inputs to multi-robot hybrid controllers which characterize and enforce collision-free interactions, take into account kinematic and geometric constraints, and are executable on actual robotic platforms. 
This is the first approach that uses braids in multi-robot motion planning to symbolically characterize and enforce rich interaction patterns, implementable on actual robotic platforms. 
In particular, we extend the notions presented in \cite{MixingBraids,MixingUnicycles}. 
The definition on planar braids and their geometric interpretations are generalized from their original form presented in \cite{MixingBraids}. 
A controller was designed for nonholonomic vehicles to track a geometric path in \cite{MixingUnicycles}. 
In this paper, we modify this controller 
to instead produce optimal trajectories that are provably safe in the collision-free sense, satisfy a set of spatio-temporal constraints, and follow desired geometric paths (Section \ref{sec:Braid_Reparameterization}). 
This also allows us to obtain tighter bounds on the amount of interaction patterns that are achievable in a space than the one found in \cite{MixingUnicycles}. 
Two new contributions to this work are found in Section \ref{sec:ArbRegions}, where these trajectory generating controllers are extended to non-rectangular regions, and Section \ref{sec:tracking_controller}, which includes development of an optimal trajectory tracking controller with formal guarantees on optimality and spatio-temporal constraint satisfaction. 

%%%%%%%%%%%%%%%%%%%%%%%%%%%%%%%%%%%%%%%%%%%%%%%%%%%%%%%%%%%%%%%%%%%%%%%%%
\subsection{Planar Braids}
\begin{figure}[tb]
\centering
\subfloat[Two robots, originally at one side of a square, map to the other.\label{fig:TwoRobotSquare}]
{\parbox[b][0.2\columnwidth][b]{0.3\columnwidth}{\centering
\includegraphics[width=0.225\columnwidth]{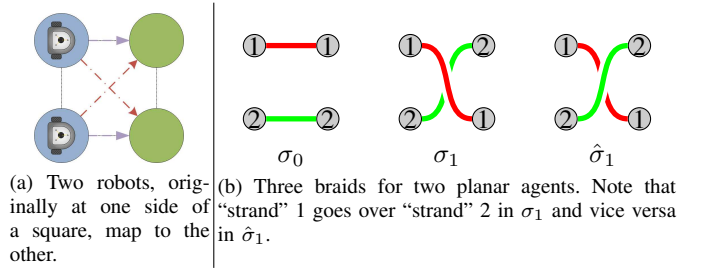}}}
\hfill\vrule\hfill
\subfloat[Three braids for two planar agents. Note that ``strand'' 1 goes over ``strand'' 2 in $\sigma_1$ and vice versa in $\hat{\sigma}_1$. \label{fig:twobraids}]{%
\parbox[b][0.2\columnwidth][b]{0.22\columnwidth}{\centering\drawcurvedbraid[5]{ 0}{2}\\
$\sigma_0$%
} 
\parbox[b][0.2\columnwidth][b]{0.22\columnwidth}{\centering\drawcurvedbraid[5]{ 1}{2}\\
$\sigma_1$%
}
\parbox[b][0.2\columnwidth][b]{0.22\columnwidth}{\centering\drawcurvedbraid[5]{-1}{2}\\
$\hat{\sigma}_1$%
}%
}
\caption[Two-robot interactions and corresponding symbolic braids]{Two-robot interactions and corresponding symbolic braids.}
\end{figure}

Consider two agents on a square, initially located at the two left vertices of the square as in Fig. \ref{fig:TwoRobotSquare}. The agents' task is to move to the two right vertices of the square. There are two ways in which these target vertices can be assigned. The first is to simply let the robots move along a straight line while the second is to have them cross paths and move to vertices diagonally across from their initial placement. 
If the robots are not to collide with each other, one agent can cross the intersection of the two paths first, and then the other (or vice versa). In the braid group, these two options correspond to different ``braids,'' and we have thus identified three planar braids for two agents, as shown in Fig. \ref{fig:twobraids}. Let us momentarily denote these three braids, $\sigma_0, \sigma_1, \hat{\sigma}_1$.

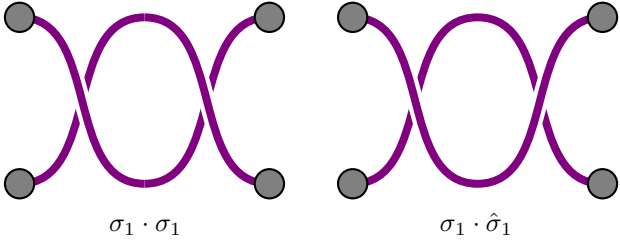
\begin{figure}[t]
\centering
\parbox{\columnwidth}{\centering%
\begin{tikzpicture}[scale=\linewidth/8cm]
%%%%%%%%%%%%%%%%%%%%%%%%%%%%%%%%%%%%%%%%%%%%%%%%%%%%%%%%%%%%%%%%%%%%%%%%%%%%%%%%%%%%%%%%%%%
\draw[name path=line 1,draw=white,ultra thick,double distance = 3pt,double=red!50!blue] (0,0) to[out=0,in=180] (1.501,2) (1.499,0) to[out=0,in=180] (3,2);
\draw[name path=line 2,draw=white,ultra thick,double distance = 3pt,double=blue!50!red] (0,2) to[out=0,in=180] (1.501,0) (1.499,2) to[out=0,in=180] (3,0);
\fill[gray] {(0,0) circle (5pt) node {}}
			{(3,0) circle (5pt) node {}}
			{(0,2) circle (5pt) node {}}
			{(3,2) circle (5pt) node {}};
\draw[thick]{(0,0) circle (5pt) node {}}
			{(3,0) circle (5pt) node {}}
			{(0,2) circle (5pt) node {}}
			{(3,2) circle (5pt) node {}};
%%%%%%%%%%%%%%%%%%%%%%%%%%%%%%%%%%%%%%%%%%%%%%%%%%%%%%%%%%%%%%%%%%%%%%%%%%%%%%%%%%%%%%%%%%%
\def\DrawingGap{1}
\draw[name path=line 3,draw=white,ultra thick,double distance = 3pt,double=red!50!blue] 
				(3+\DrawingGap,0) to[out=0,in=180]  (4.5+\DrawingGap,2) to[out=0,in=180]  (6+\DrawingGap,0);
\draw[name path=line 4,draw=white,ultra thick,double distance = 3pt,double=blue!50!red] 
				(3+\DrawingGap,2) to[out=0,in=180]  (4.5+\DrawingGap,0) to[out=0,in=180]  (6+\DrawingGap,2);
\fill[gray] {(3+\DrawingGap,0) circle (5pt) node {}}
			{(6+\DrawingGap,0) circle (5pt) node {}}
			{(3+\DrawingGap,2) circle (5pt) node {}}
			{(6+\DrawingGap,2) circle (5pt) node {}};
\draw[thick]{(3+\DrawingGap,0) circle (5pt) node {}}
			{(6+\DrawingGap,0) circle (5pt) node {}}
			{(3+\DrawingGap,2) circle (5pt) node {}}
			{(6+\DrawingGap,2) circle (5pt) node {}};
%%%%%%%%%%%%%%%%%%%%%%%%%%%%%%%%%%%%%%%%%%%%%%%%%%%%%%%%%%%%%%%%%%%%%%%%%%%%%%%%%%%%%%%%%%%
\end{tikzpicture}%
}\\[0.5em]
$\sigma_1\cdot\sigma_1$\hspace{9.75em}$\sigma_1\cdot\hat{\sigma}_1$
\caption[Concatenated braids generators]{Two concatenated braids $\sigma_1\cdot\sigma_1$ and $\sigma_1\cdot\hat{\sigma}_1$. The latter of these two braids is the same as the identity braid $\sigma_0$.}
\label{fig:concbraid}
\end{figure}

Now, given these three braids, we can concatenate them together to form other braids, as seen in Fig. \ref{fig:concbraid}. The left braid is given by $\sigma_1\cdot\sigma_1$ and the right braid is $\sigma_1\cdot\hat{\sigma}_1$. In the braid group, what really matters is not the geometric layout of the paths, but how the paths wrap around each other. As can be seen, if we were to ``pull" the right corners in $\sigma_1\cdot\sigma_1$, the ``strands'' would get ``tangled" in the middle, while a ``stretched-out" $\sigma_1\cdot\hat{\sigma}_1$ is simply $\sigma_0$. Thus we let $\sigma_0$ be the identity braid, such that $\sigma_1$ and $\hat{\sigma}_1$ are each others' inverses in the sense that
\begin{equation*}
\sigma_1\cdot\hat{\sigma}_1=\hat{\sigma}_1\cdot\sigma_1=\sigma_0.
\end{equation*}
In fact, every braid has an inverse and, as such, the set of braids (together with the concatenation operation) is indeed a group. And, as $\sigma_1^{-1}=\hat{\sigma}_1$ (and $\sigma_1^{-1}=\hat{\sigma}_1$),  $\sigma_0$ and $\sigma_1$ (or $\sigma_0$ and $\hat{\sigma}_1$ for that matter) are the so-called generator braids for this group in that all planar braids can be written as concatenations of these two braids and their inverses \cite{braid1,braid2}. 

This notion of generator braids can be extended to the case when there are $N\geq2$, % rather than two agents, 
with the only difference being that we will have $N$ generators rather than just two, i.e.,  let $\sigma_0$ be the trivial generator with no interactions and $\sigma_k$ denote the interactions between agents $k$ and $k+1$, $k=1,\ldots,N-1$. 
If we let $\Sigma_N$ be the set of all planar generator braids over $N$ agents, this set will serve as the alphabet over which braid strings (themselves braids) are produced from, and we let $\Sigma_N^M$ denote the set of all braids of length $M$ (i.e., braids composed of $M$ generators) over $N$ agents.

%%%%%%%%%%%%%%%%%%%%%%%%%%%%%%%%%%%%%%%%%%%%%%%%%%%%%%%%%%%%%%%%%%%%%%%%%%%%%
\subsection{A Geometric Interpretation}\label{sec:GeomInterp}
Although heavily inspired by geometry in \cite{braid1}, the braid group is not concerned with the actual geometry of the braid strands. For the sake of describing the robot motion plans,
we will associate geometric paths with the different braids. First of all, we assume that the braid is geometrically located in a rectangular area of height $h$ and length $\ell$ no matter how long the braid string is. Using the particular two-agent braids discussed in the previous paragraphs, we assume that the two agents are initially located at the points $(0,0)$ and $(0,h)$, while the final locations are at $(\ell,0)$ and $(\ell,h)$. If the total braid results from the use of one single generator braid $\sigma_0$, or $\sigma_1$ then no additional points are needed. However, if the braid has length 2, then we also need to introduce intermediary ``half-way" points $(\ell/2,0)$ and $(\ell/2,h)$. As such, we let ${\mathcal{P}}_2^q=\left\{\left(\frac{q}{M}\ell,0\right),\left(\frac{q}{M}\ell,h\right)\right\}$ be a set of uniformly spaced%
\footnote{Unless otherwise stated, the rest of this document we will assume these sets of intermediary points are uniformly spaced in $h$ and $\ell$, but the notions presented here extend to points which are not, as the application demands. More on this in Section \ref{sec:ArbRegions}.}
 positions for these intermediary points, where the subscript 2 denotes the two-agent case, $M$ is the length of the braid to be executed, and $q=0,1,\ldots,M$. 

Using this notation, we can refer back to Fig. \ref{fig:TwoRobotSquare} and say that each of the two generator braids correspond to an assignment, i.e., a bijective map, between ${\mathcal{P}}_2^{0}$ and ${\mathcal{P}}_2^1$, and we use the following notation to denote this fact
\begin{equation*}
\sigma_i:{\mathcal{P}}^{0}_2 \rightarrow_b{\mathcal{P}}^{1}_2,~~i = 0,1,
\end{equation*}
where $\rightarrow_b$ denotes ``bijection". Note that this is a slight abuse of notation in that $\sigma_i$ now denotes both an element in the braid group as well as a map -- this distinction, however, should be entirely clear from the context. Further, we will refer to these points which agents are bijectively mapped to and from as \emph{braid points}.

If we generalize this to $N\geq 2$ agents and let $\sigma$ denote a string of generators of length $M\geq2$, i.e., $\sigma\in\Sigma_N^M$, we will use the notation
\begin{equation*}
\sigma(k):{\mathcal{P}}^{(k-1)}_N\rightarrow_b{\mathcal{P}}^{k}_N, \quad k=1,\ldots,M,
\end{equation*}
where $\sigma\left(k\right)$ is the $k^\mathrm{th}$ braid\footnote{Note that a ``braid'' here refers to a member of the braid group, e.g., a single generator (viz., a single bijective map) or a concatenation of several generators (viz., a composition of bijective maps).} in the string $\sigma$ and
\begin{multline*}
\mathcal{P}^q_N=\left\{\left(\tfrac{q}{M}\ell,0\right),\left(\tfrac{q}{M}\ell,\tfrac{1}{(N-1)}h\right),\right.\\
\left.\left(\tfrac{q}{M}\ell,\tfrac{2}{(N-1)}h\right),\ldots,\left(\tfrac{q}{M}\ell,h\right)\right\},
\end{multline*}
as shown for the three-agent case in Fig. \ref{fig:bijection}.

\begin{figure}[tb]
\centering
$\overset{\mbox{$\sigma(1)$}}{\scalebox{8}[2.5]{$\curvearrowright$}}$
\hspace{0.1em}
$\overset{\mbox{$\sigma(2)$}}{\scalebox{8}[2.5]{$\curvearrowright$}}$%\quad\ 
\\[0.5em]
\scalebox{0.9}{
\parbox[c][][c]{\columnwidth}{\centering$\underbrace{\drawbraid*[5][0.6][5]{2,1}{3}}_{\mbox{$\ell$}}$\raisebox{3.0em}{\scalebox{3}[13]{\}}}}\rlap{\hspace{-1.5em}\raisebox{0.5em}{$h$}}}\\
\small
$\mathcal{P}^{0}_3=\left\{\left(0,0\right),\left(0,0.5h\right),\left(0,h\right)\right\}$
\hfill%\hspace{1em}
$\mathcal{P}^{2}_3=\left\{\left(\ell,0\right),\left(\ell,0.5h\right),\left(\ell,h\right)\right\}$\\
$\mathcal{P}^{1}_3=\left\{\left(0.5\ell,0\right),\left(0.5\ell,0.5h\right),\left(0.5\ell,h\right)\right\}$
\caption[Geometric interpretation of braid string]{The geometric interpretation of braid string $\sigma = \sigma_2\cdot\sigma_1$ for the three-agent case. In this example, $\sigma(1)=\sigma_2$ and maps $\mathcal{P}^0_3$ to $\mathcal{P}^{1}_3$, while $\sigma(2)=\sigma_1$ and maps $\mathcal{P}^{1}_3$ to $\mathcal{P}^{2}_3$.}
\label{fig:bijection}
\end{figure}

We moreover use the notation $\xi(i,j)\in\R^2$ to denote the point agent $j$ should go to at step $i,~i=1,\ldots,M$. We use the convention that $\xi(0,j)=(0,(j-1)h/(N-1))$,  $j=1,2,\ldots,N$, to denote agent $j$'s initial position.  In other words,
\begin{align*}
\xi(1,j)&=\sigma(1)\langle\xi(0,j)\rangle,\\
\xi(2,j)&=\sigma(2)\langle\xi(1,j)\rangle=\sigma(2)\circ\sigma(1)\langle\xi(0,j)\rangle,
\intertext{or more generally,}
\xi(i,j)&=\sigma(i)\langle\xi(i-1,j)\rangle\\ 
&=\sigma(i)\circ\sigma(i-1)\circ\cdots\circ\sigma(1)\langle\xi(0,j)\rangle,
\end{align*}
where we use the $\langle\cdot\rangle$ notation to denote the argument to $\sigma(i)$ and $\circ$ to denote composition. This construction is also illustrated in Fig. \ref{fig:bijection} for the three-agent case.

The geometric interpretation we will make of the planar braids is that the mobile agents that are to execute them must traverse through these braid points. They must moreover do so in an orderly and safe manner, which will be the topic of the next section.
%%%%%%%%%%%%%%%%%%%%%%%%%%%%%%%%%%%%%%%%%%%%%

\section{Executing Braids}\label{sec:braid_controllers}
Given a collection of $N$ agents with dynamics
\begin{align*}
&\dot x_j=f(x_j,u_j),
\end{align*}
%\intertext
{and planar output}
\begin{align*}
&y_j=h(x_j)\in\R^2,~~j=1,\ldots,N,
\end{align*}
then it is of interest to have these agents execute a braid $\sigma\in\Sigma_N^M$. We now define what it means for this braid to be executed. 

Given an input braid string $\sigma$, what each individual agent should do is ``hit" the intermediary braid points $\xi(i,j)$ at specified time instances. We let $T$ denote the time it should take for the entire string to be executed. As such, the first condition for a multi-agent motion to be feasible with respect to the braid is the following:

\begin{definition}[Braid-Point Feasibility]\ \\
A multi-robot trajectory is \emph{braid-point feasible} if
\begin{equation*}
y_j(t_i)=\xi(i,j),~i=0,\ldots,M,~j=1,\ldots,N.
\end{equation*}
where the $t_i$'s form a partition of a given time window $[0,T]$, i.e., 
\begin{equation*}
t_0=0<t_1<\cdots<t_i<\cdots<t_M= T.\tag*{$\diamond$}
\end{equation*}
\end{definition}

On top of braid-point feasibility, we also insist on the robots not colliding during the maneuvers. To a certain degree, this condition is what restricts the level of \emph{mixing} that is possible, i.e., since the braid is constrained in a rectangle of fixed height and width, what length strings the multi-robot system can execute while maintaining a desired level of safety separation.

\begin{definition}[Collision-Free]\ \\
A multi-robot trajectory is \emph{collision-free} if
\begin{equation*}
\|y_i(t)-y_j(t)\|\geq\delta_{ij},\quad\forall i\neq j,\quad t\in[0,T],
\end{equation*}
where $\delta_{ij}>0$ is the desired level of safety separation between agents $i$ and $j$.\hfill$\diamond$
\end{definition}

For convenience, we will refer to $\bar\delta = \max\delta_{ij}$ as the \emph{maximum safety separation} such that no agent collides. This will come up in Theorem \ref{thm:MixingLimit}.
We are missing a notion to describe what the multi-robot mixing problem is, that is, what constitutes a \emph{braid controller}.
\begin{definition}[Braid Controller]\ \\
A multi-robot controller is a \emph{braid controller} if the resulting trajectories are both braid-point feasible and collision free, for all collision-free initial conditions such that \begin{equation*}
y_j(0)=\xi(0,j),~j=1,\ldots,N. \tag*{$\diamond$}
\end{equation*}
\end{definition}

As a final notion, we are interested in how much mixing a particular system can support. 
\begin{definition}[Mixing Limit]\ \\
The \emph{mixing limit} $M^\star$ is the largest integer $M$ such that there exists a braid controller for every string in $\Sigma_N^M$.\hfill$\diamond$
\end{definition}

Whenever two strands of the braid associated with a given braid string cross, the two associated agents will have to interact. The mixing limit therefore serves as an input-independent bound on how much mixing is achievable for a  given team of agents operating in a given environment.
The mixing limit is in general quite hard to compute, it needs to consider every permutation of strings of varying lengths up to some number, the geometry assigned to each string and is dependent on the kinematical response of the multi-robot system.
However, under certain assumptions 
it is possible to find bounds on $M^\star$ for a given braid controller.

%%%%%%%%%%%%%%%%%%%%%%%%%%%%%%%%%%%%%%%%%%%%%%%%%%%%%%%%%%%%%%%%%%%%%%%%%%%
\subsection{Braid Controllers: Stop-Go-Stop Hybrid Strategy}
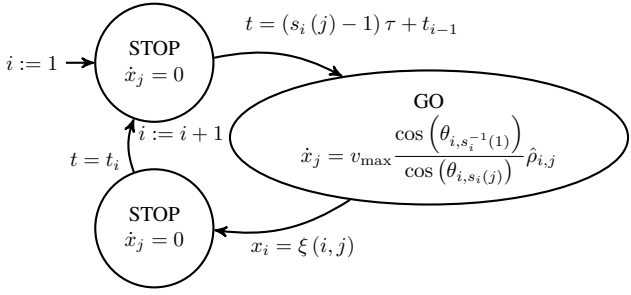
\begin{figure}[t]
\centering
\tikzset{elliptic state/.style={draw,ellipse}}%
\begin{tikzpicture}[->,>=stealth',%
					initial text={$i:=1$},thick,%
					scale=0.8,%
					every node/.style={transform shape}]

 % State: STOP_1
 \node[initial left,state,anchor=west] (STOP_1)
 {\begin{tabular}{c}
  STOP \\ $\dot{x}_j = 0$
 \end{tabular}};

 % State: GO
 \node[elliptic state,%/.style = {draw,ellipse},           % layout (defined above)
  below right of=STOP_1,% Position is to the right of STOP_1
  node distance=5em,    % distance to STOP_1
  inner sep=0pt, minimum size=0pt,
  anchor=west] (GO) % posistion relative to the center of the 'box'
 {%
 \begin{tabular}{c}     % content
  GO \\ $\dot{x}_j = v_\mathrm{max}\dfrac{\cos\left(\theta_{i,s_i^{-1}(1)}\right)}{\cos\left(\theta_{i,s_i(j)}\right)}\hat{\rho}_{i,j}$
 \end{tabular}
 };

 % State: STOP_2
 \node[state,                % layout (defined above)
  below of=STOP_1, 			 % Position is to the bottom of GO
  node distance=5.0em,       % distance to GO
  anchor=north] (STOP_2) 	 % posistion relative to the center of the 'box'
 {%
 \begin{tabular}{c}          % content
  STOP \\ $\dot{x}_j = 0$
 \end{tabular}
 };

 \path (STOP_1) edge[bend left=20] 
		node[anchor=west,above right,xshift=-2.0em,yshift=0.5em]{$t=\left(s_i\left(j\right)-1\right)\tau+t_{i-1}$}
 	   (GO)
       (GO) edge[bend left=20] 
        node[anchor=east,below right,xshift=-2.0em]{$x_i = \xi\left(i,j\right)$} 
       (STOP_2)
       (STOP_2) edge[bend left=20] 
        node[anchor=west,below left]{$t=t_i$} 
        node[anchor=north, right,yshift=+0.5em,xshift=-0.5em]{\begin{tabular}{l}$i:=i+1$\end{tabular}} 
       (STOP_1)
       ;

\end{tikzpicture}
\caption[Hybrid STOP-GO-STOP braid controller]%
{Hybrid STOP-GO-STOP braid controller.\label{fig:STOP_GO_STOP}}
\end{figure}

Our first attempt at executing braids will be a hybrid control strategy called the \emph{Stop-Go-Stop}. We will assume that agent dynamics are given by single integrator dynamics, i.e.,
%\begin{align*}
$\dot{x}_j = u_j\in\R^2$
%\end{align*}
with $y_j = x_j$, $j=1,\ldots,N$. Further, for practical considerations, assume that there is a cap on the maximum velocity achievable by the agents, i.e., $\|u_j\| \in \left[0,v_\mathrm{max}\right]$.
The idea is then at each braid step to ``send'' agents off straight to their next braid point in order of the distance they'll need to travel, waiting just long enough to avoid collisions before sending an agent off.
To that end, we define $s_i:\left\{1,\ldots,N\right\}\to\left\{1,\ldots,N\right\}$ to be a bijective mapping that denotes the farthest distance ordering at step $i$, that is
\begin{multline*}
s_i(p)<s_i(q)\qquad\Longrightarrow\qquad \\
\norm{\xi(i,p)-\xi(i-1,p)}\geq\norm{\xi(i,q)-\xi(i-1,q)}
\end{multline*}
where ties are arbitrarily broken such that $s_i$ remains a bijection. We let the agents heading angle be given by 
%\begin{align*}
$\theta_{i,j} = \tan^{-1}\left(\frac{\left(\xi(i,j)-\xi(i-1,j)\right)_2}{\left(\xi(i,j)-\xi(i-1,j)\right)_1}\right)$
%\end{align*}
where the subscript indicates the first or second component, and the unit heading vector be $\hat{\rho}_{i,j} = \left[\cos(\theta_{i,j}),\sin(\theta_{i,j})\right]^T$. Lastly, the time the agents will wait before entering \emph{GO} mode will be given by their ordering as $(s_i(j)-1)\tau$ where
\begin{align*}
\tau = \frac{\delta}{v_\mathrm{max}\cos(\theta^*)}
\end{align*}
is the time required to be $\delta$ apart horizontally\footnote{Since we are interested in the mixing limit, the analysis is done with the horizontal direction being the limiting factor for safe execution of the braids.}, with $\cos(\theta^\star) = \frac{\ell/M}{\sqrt{\ell^2/M^2+h^2}}$ being the maximum horizontal distance an agent could travel given $\sigma\in\Sigma_N^M$. To ensure that the agents do not overtake the first agent (horizontally), the speed of the agents should be scaled by the velocity of the first agent, i.e., $u_j = v_\mathrm{max}\cos(\theta_{i,s_i^{-1}(1)})/\cos(\theta_{i,s_i(j)})$. The hybrid automaton describing the stop-go-stop controller is given in Fig. \ref{fig:STOP_GO_STOP}.

\begin{theorem}{\cite{MixingBraids}}\label{thm:STOP_GO_TOP}
The STOP-GO-STOP controller in Fig. \ref{fig:STOP_GO_STOP} is a braid controller if the braid points themselves are sufficiently separated and
\begin{multline*}
\cos(\theta^\star)v_{\max}\left(\min_i\left(t_i-t_{i-1}\right)-(N-1)\tau\right)\\
\geq\sqrt{\ell^2/M^2+h^2}.
\end{multline*}
\end{theorem}
{\it Proof:}\newline\noindent
The STOP-GO-STOP controller ensures that the agents are never within $\delta$ of each other by virtue of the fact that they have to wait until they are indeed at least that far apart (horizontally) before entering GO mode.
As such, the trajectories are collision-free. 

What remains to show is that they are also braid-point feasible. Consider the agent that has to wait the longest before it can move, i.e., the agent that has to wait a total of $(N-1)\tau$, and at its worst has $\min_i\left(t_i-t_{i-1}\right)-(N-1)\tau$ time left to reach the next braid point. In other words, we need that the distance traveled in that amount of time at the speed $v_{\max}\cos(\theta_{i,s_i^{-1}(1)})/\cos(\theta_{i,s_i(j)})$
is greater than the distance required. But, we note that
\begin{equation*}
v_{\max}\cos(\theta_{i,s_i^{-1}(1)})/\cos(\theta_{i,s_i(j)})\geq v_{\max}\cos(\theta^\star)
\end{equation*}
and, as we are only looking for a bound, we assume that we use this lower speed and that the distance required to travel is the largest distance possible (which it really is not). In other words, we need
\begin{multline*}
\cos(\theta^\star)v_{\max}\left(\min_i\left(t_i-t_{i-1}\right)-(N-1)\tau\right)\\
\geq\sqrt{\ell^2/M^2+h^2},
\end{multline*}
and the proof follows.\qed

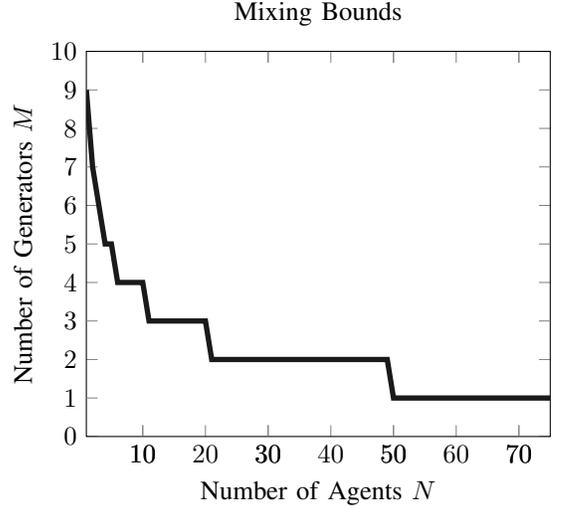
\begin{figure}
\centering
\begin{tikzpicture}
\begin{axis}[scale=0.9,%
			 compat=newest,
			 xlabel={Number of Agents $N$},%
			 ylabel={Number of Generators $M$},%
			 ymin = 0, ymax = 10,%
			 xmin = 1, xmax = 75,%
			 extra x ticks = {10,30,...,70},%
			 extra y ticks = {1,3,...,9},%
			 title = {Mixing Bounds},%
			 ]%
\addplot[no marks,black!90!white,line width=2pt,solid] coordinates {%
		(1,9)
		(2,7)
		(4,5)
		(5,5)
		(6,4)
		(10,4)
		(11,3)
		(20,3)
		(21,2)
		(49,2)
		(50,1)
		(75,1)
		};%
\end{axis}
\end{tikzpicture}
\caption{Lower bound on the mixing limit using the Stop-Go-Stop braid controller.}
\label{fig:mixingbound}
\end{figure}

Note that Theorem \ref{thm:STOP_GO_TOP} implicitly provides a lower bound on the mixing limit as long as the agents' paths are straight lines. 
In Fig. \ref{fig:mixingbound} we can see the mixing limit as a function of the number of agents in the team for parameters $v_{\max}=5,~T=20,~\ell=5,~h=10$, and $\Delta=0.2$.
The problem with this strategy is that it does not allow for more general geometric paths (which could be interpreted as feasible trajectories under a given robot dynamical model), nor does it ensure that agents get within a specified distance from each other (which could be necessary when collaboration is required). The next section we present 
a new strategy called \emph{braid reparameterization} which will explicitly consider more general strand geometries, and allow us to obtain analytical bounds on the mixing limit.

%%%%%%%%%%%%%%%%%%%%%%%%%%%%%%%%%%%%%%%%%%%%%%%%%%%%%%%%%%%%%%%%%%%%%%%%%%%%%%

\subsection{Braid Controller: Braid Reparameterization}\label{sec:Braid_Reparameterization}

We are now seeking a strategy that will allow us to follow a given geometry while achieving a mixing strategy encoded as a braid string $\sigma\in\Sigma_N^M$. Further, we wish to enforce inter-agent interaction as dictated in $\sigma$ by having agents get as close as the safety separation $\delta_{jk}$. Before moving forward, consider the following lemma. 
\begin{lemma}\label{lem:simult_braids}
If at any braid step, the generators in the braid substring $\sigma(k)=\sigma\subseteq\Sigma_N^m$, $m\leq M$, 
have indices that are two or more apart,
then any agent interacts with at most one other agent at this step.
%braid step $k$.
\end{lemma}

\noindent
\textit{Proof:}\newline\noindent
Let $\mathcal{B}_N^k\in\R^{N\times2}$ be a matrix that contains the set of braid points at time $k$ such that $\mathcal{B}_N^0=[\xi(0,1) \cdots \xi(0,N)]^T$.
Consider the two-generator concatenation $\sigma_i\cdot\sigma_j$.
As a bijective map, if $\sigma_i=\sigma_0$, then the agents do not interact, and the agent in braid point position $[\mathcal{B}_N^{k-1}]_n$ gets mapped to braid point position $[\mathcal{B}_N^{k}]_n$, where $[\mathcal{B}]_n$ corresponds to the $n^\mathrm{th}$ row of $\mathcal{B}$, $n=1,\ldots,N$. 
If $\sigma_i\neq\sigma_0$, then $\sigma_i$ will swap the position of the two agents occupying the braid point positions $i$ and $i+1$ at step $k-1$, 
i.e., it maps the agent occupying $[\mathcal{B}_N^{k-1}]_i$ to $[\mathcal{B}_N^{k}]_{i+1}$ and the agent occupying $[\mathcal{B}_N^{k-1}]_{i+1}$ to $[\mathcal{B}_N^{k}]_{i}$.
Similarly, $\sigma_j$ swaps the position of the agents occupying the braid point positions $j$ and $j+1$. 

The two agents in positions $i$ and $i+1$ at $k-1$ would only interact with the two agents in positions $j$ and $j+1$ at $k-1$
if $i$ (or $i+1$ for that matter) is equal to either $j$ or $j+1$.
But if we let $\abs{i-j}\geq 2$, then we get that
\begin{align*}
\abs{i-j}\geq 2~
&\Leftrightarrow~2\leq i-j~\mathrm{or} ~2\leq j-i\\
&\Rightarrow~j<j+1<j+2\leq i<i+1~\\
&~\mathrm{or} ~~\,i<i+1<i+2\leq j<j+1\\
%j-2\leq i\leq j+2\\
%&\Leftrightarrow~j-1< i< j+1\\%[-1.75em]
&\Rightarrow%~j< i+1< j+2~\Leftrightarrow
%~j-1< i< j+1\\
{\left\{
\begin{array}{l}
%j-1\neq i-1\\
%j-1\neq i\\
%j-1\neq i+1\\
%j\neq i-1\\
%j+1\neq i-1\\
i\neq j\\
i\neq j+1\\
i+1\neq j\\
i+1\neq j+1,
\end{array}\right.}
%&\Rightarrow~j-1\neq i-1\\
%&\Rightarrow~j-1\neq i\\
%&\Rightarrow~j-1\neq i+1\\
%&\Rightarrow~j\neq i-1\\
%&\Rightarrow~j\neq i\\
%&\Rightarrow~j\neq i+1\\
%&\Rightarrow~j+1\neq i-1\\
%&\Rightarrow~j+1\neq i\\
%&\Rightarrow~j+1\neq i+1,
\end{align*}
and as such these two-generator concatenation maps the agents from one set of braid points to the next with at most two interaction between at most two agents per interaction.

One of the two braid group relations \cite{braid1,braid2} tells us that if $\abs{i-j}\geq2$ then $\sigma_i\cdot\sigma_j = \sigma_j\cdot\sigma_i$.
More generally, if we let $$h:\{1,\ldots,m\}\to\{1,\ldots,N-1\}$$ be a surjective map such that
$\abs{h(i)-h(j)}\geq 2$ for all $i,j\in\{1,\ldots,m\}$ with $i \neq j$, then for \emph{any} bijective map $g:\{1,\ldots,m\}\to\{1,\ldots,m\}$ we have that
\begin{multline*}
\sigma(k) = \sigma_{h(1)}\cdot\sigma_{h(2)}\cdot\cdots\cdot\sigma_{h(m)}\\
=\sigma_{h(g(1))}\cdot\sigma_{h(g(2))}\cdot\cdots\cdot\sigma_{h(g(m))}.
\end{multline*}
As such, 
the braid generators can be rearranged to obtain any permutation of two-generator concatenations from generators in $\sigma(k)$. Since for all permutations of two-generator concatenations we will have indices that are two or more apart, these will map the agents at braid step $k$ from the set of braid points $\mathcal{B}^{k-1}_N$ to the next set of braid points $\mathcal{B}^{k}_N$ with at most $m$ interactions total, and at most one interaction per agent.
\hfill\qed

If the geometric interpretation of the braid string is restricted to the case of only pairwise interactions at every braid step, then it is possible to devise a hybrid strategy with the desired properties, which we can then compose together to achieve the desired interaction patterns. As such, we
will restrict the geometric interpretation of braid strings 
to those satisfying pairwise interactions as in Lemma \ref{lem:simult_braids}.

\begin{restr}\label{res:simult_braids}
As a bijective map from one set of braid points to another, the braid $\sigma(k)$
will only contain generators whose indices 
are two or more apart, 
i.e., for some
%any $k$ and 
$h:\mathbb{N}\to\{0,\ldots,N-1\}$
\begin{align*}
\sigma(k) = \sigma_{h(1)} \cdot \sigma_{h(2)} \cdot \cdots, ~~
 \mathrm{where}~\abs{h(i)-h(j)}\geq 2~\forall i\neq j
\end{align*}
\end{restr}

\begin{figure}[t]
\centering
\begin{tikzpicture}[scale=\linewidth/10cm]
%%%%%%%%%%%%%%%%%%%%%%%%%%%%%%%%%%%%%%%%%%%%%%%%%%%%%%%%%%%%%%%%%%%%%%%%%%%%%%%%%%%%%%%%%%%% SIGMA(1) = SIGMA_1*SIGMA_3*SIGMA_2 %%%%%%%%%%%%%%%%%%%%%%%%
\pgfmathsetmacro{\nodegap}{3}
\pgfmathsetmacro{\horzgap}{3/2}
\pgfmathsetmacro{\vertgap}{-6}
\pgfmathsetmacro{\figgap}{0}
\coordinate (D1) at ($(0,0)+(\figgap,0)$); 
\coordinate (D2) at ($(D1)+(0.5*\nodegap,0)$);
\coordinate (D3) at ($(D1)+(\nodegap,0)$);
\coordinate (C1) at ($(0,1)+(\figgap,0)$); 
\coordinate (C2) at ($(C1)+(0.5*\nodegap,0)$);
\coordinate (C3) at ($(C1)+(\nodegap,0)$);
\coordinate (B1) at ($(0,2)+(\figgap,0)$);
\coordinate (B2) at ($(B1)+(0.5*\nodegap,0)$);
\coordinate (B3) at ($(B1)+(\nodegap,0)$);
\coordinate (A1) at ($(0,3)+(\figgap,0)$);
\coordinate (A2) at ($(A1)+(0.5*\nodegap,0)$);
\coordinate (A3) at ($(A1)+(\nodegap,0)$);

\draw[name path=line 4,draw=white,ultra thick,
		double distance = 3pt,double=blue!50!red] 
		(D1) -- (B3);
\draw[name path=line 3,draw=white,ultra thick,
		double distance = 3pt,double=blue!50!red] 
		(C1) -- (D3);
\draw[name path=line 2,draw=white,ultra thick,
		double distance = 3pt,double=blue!50!red] 
		(B1) -- (A3);
\draw[name path=line 1,draw=white,ultra thick,
		double distance = 3pt,double=red!50!blue] 
		(A1) -- (C3);
\fill[gray] {(A1) circle (5pt) node {}}
			{(A3) circle (5pt) node {}}
			{(B1) circle (5pt) node {}}
			{(B3) circle (5pt) node {}}
			{(C1) circle (5pt) node {}}
			{(C3) circle (5pt) node {}}
			{(D1) circle (5pt) node {}}
			{(D3) circle (5pt) node {}};
\draw[thick]{(A1) circle (5pt) node {}}
			{(A3) circle (5pt) node {}}
			{(B1) circle (5pt) node {}}
			{(B3) circle (5pt) node {}}
			{(C1) circle (5pt) node {}}
			{(C3) circle (5pt) node {}}
			{(D1) circle (5pt) node {}}
			{(D3) circle (5pt) node {}};
\node[label={$\genfrac{}{}{0pt}{}{\displaystyle \{\sigma_1\cdot\sigma_3\cdot\sigma_2\}}{\displaystyle\ \qquad=\{\sigma_3\cdot\sigma_1\cdot\sigma_2\}}$}, below, yshift = -3.5em] at (D2) {};
\node[label={(a)}, below, yshift = -5em] at (D2) {};
%%%%%%%%%%%%%%%%%%%%%%%%%%%%%%%%%%%%%%%%%%%%%%%%%%%%%%%%%%%%%%%%%%%%%%%%%%%%%%%%%%%%%%%%% SIGMA(1) = SIGMA_1*SIGMA_3, SIGMA(2) = SIGMA_2 %%%%%%%%%%%%%%%
\pgfmathsetmacro{\figgap}{\horzgap*\nodegap}
\coordinate (D1) at ($(0,0)+(\figgap,0)$); 
\coordinate (D2) at ($(D1)+(0.5*\nodegap,0)$);
\coordinate (D3) at ($(D1)+(\nodegap,0)$);
\coordinate (C1) at ($(0,1)+(\figgap,0)$); 
\coordinate (C2) at ($(C1)+(0.5*\nodegap,0)$);
\coordinate (C3) at ($(C1)+(\nodegap,0)$);
\coordinate (B1) at ($(0,2)+(\figgap,0)$);
\coordinate (B2) at ($(B1)+(0.5*\nodegap,0)$);
\coordinate (B3) at ($(B1)+(\nodegap,0)$);
\coordinate (A1) at ($(0,3)+(\figgap,0)$);
\coordinate (A2) at ($(A1)+(0.5*\nodegap,0)$);
\coordinate (A3) at ($(A1)+(\nodegap,0)$);

\draw[name path=line 4,draw=white,ultra thick,
		double distance = 3pt,double=blue!50!red] 
		(D1) -- (C2) -- (B3);
\draw[name path=line 3,draw=white,ultra thick,
		double distance = 3pt,double=blue!50!red] 
		(C1) -- (D2) -- (D3);
\draw[name path=line 2,draw=white,ultra thick,
		double distance = 3pt,double=blue!50!red] 
		(B1) -- (A2) -- (A3);
\draw[name path=line 1,draw=white,ultra thick,
		double distance = 3pt,double=red!50!blue] 
		(A1) -- (B2) -- (C3);
\fill[gray] {(A1) circle (5pt) node {}}
			{(A2) circle (5pt) node {}}
			{(A3) circle (5pt) node {}}
			{(B1) circle (5pt) node {}}
			{(B2) circle (5pt) node {}}
			{(B3) circle (5pt) node {}}
			{(C1) circle (5pt) node {}}
			{(C2) circle (5pt) node {}}
			{(C3) circle (5pt) node {}}
			{(D1) circle (5pt) node {}}
			{(D2) circle (5pt) node {}}
			{(D3) circle (5pt) node {}};
\draw[thick]{(A1) circle (5pt) node {}}
			{(A2) circle (5pt) node {}}
			{(A3) circle (5pt) node {}}
			{(B1) circle (5pt) node {}}
			{(B2) circle (5pt) node {}}
			{(B3) circle (5pt) node {}}
			{(C1) circle (5pt) node {}}
			{(C2) circle (5pt) node {}}
			{(C3) circle (5pt) node {}}
			{(D1) circle (5pt) node {}}
			{(D2) circle (5pt) node {}}
			{(D3) circle (5pt) node {}};
\node[label={$\genfrac{}{}{0pt}{}{\displaystyle \{\sigma_1\cdot\sigma_3\}\cdot\sigma_2}{\displaystyle\ \qquad=\{\sigma_3\cdot\sigma_1\}\cdot\sigma_2}$}, below, yshift = -3.5em] at (D2) {};
\node[label={(b)}, below, yshift = -5em] at (D2) {};
%%%%%%%%%%%%%%%%%%%%%%%%%%%%%%%%%%%%%%%%%%%%%%%%%%%%%%%%%%%%%%%%%%%%%%%%%%
\end{tikzpicture}\\
\begin{tikzpicture}[scale=\linewidth/10cm]
%%%%%%%%%%%%%%%%%%%%%%%%%%%%%%%%%%%%%%%%%%%%%%%%%%%%%%%%%%%%%%%%%%%%%%%%%%%%%%%%%%%%%%%%%%%% SIGMA(1) = SIGMA_1*SIGMA_3*SIGMA_2 %%%%%%%%%%%%%%%%%%%%%%%%
\pgfmathsetmacro{\nodegap}{3}
\pgfmathsetmacro{\horzgap}{3/2}
\pgfmathsetmacro{\vertgap}{-6}
\pgfmathsetmacro{\figgap}{0}
%%%%%%%%%%%%%%%%%%%%%%%%%%%%%%%%%%%%%%%%%%%%%%%%%%%%%%%%%%%%%%%%%%%%%%%%%%%%%%%%%%%%%%%%%%%%%%%%%%%%%%%%%%%%%%%%%%%%%%%%%%%%%%%%%%%%%%%%%%%%%%%%%%%%%%%%%%%%%%% SIGMA(1) = SIGMA_1, SIGMA(2) = SIGMA_3, SIGMA(3) = SIGMA_1 %%%%%%%%%
\pgfmathsetmacro{\figgap}{0}
\coordinate (D1) at ($(0,0)+(\figgap,\vertgap)$); 
\coordinate (D2) at ($(D1)+(1/3*\nodegap,0)$);
\coordinate (D3) at ($(D1)+(2/3*\nodegap,0)$);
\coordinate (D_Mid) at ($0.5*(D2)+0.5*(D3)$);
\coordinate (D4) at ($(D1)+(\nodegap,0)$);
\coordinate (C1) at ($(0,1)+(\figgap,\vertgap)$); 
\coordinate (C2) at ($(C1)+(1/3*\nodegap,0)$);
\coordinate (C3) at ($(C1)+(2/3*\nodegap,0)$);
\coordinate (C4) at ($(C1)+(\nodegap,0)$);
\coordinate (B1) at ($(0,2)+(\figgap,\vertgap)$);
\coordinate (B2) at ($(B1)+(1/3*\nodegap,0)$);
\coordinate (B3) at ($(B1)+(2/3*\nodegap,0)$);
\coordinate (B4) at ($(B1)+(\nodegap,0)$);
\coordinate (A1) at ($(0,3)+(\figgap,\vertgap)$);
\coordinate (A2) at ($(A1)+(1/3*\nodegap,0)$);
\coordinate (A3) at ($(A1)+(2/3*\nodegap,0)$);
\coordinate (A4) at ($(A1)+(\nodegap,0)$);

\draw[name path=line 4,draw=white,ultra thick,
		double distance = 3pt,double=blue!50!red] 
		(D1) -- (D2) -- (C3) -- (B4);
\draw[name path=line 3,draw=white,ultra thick,
		double distance = 3pt,double=blue!50!red] 
		(C1) -- (C2) -- (D3) -- (D4);
\draw[name path=line 2,draw=white,ultra thick,
		double distance = 3pt,double=blue!50!red] 
		(B1) -- (A2) -- (A3) -- (A4);
\draw[name path=line 1,draw=white,ultra thick,
		double distance = 3pt,double=red!50!blue] 
		(A1) -- (B2) -- (B3) -- (C4);
\fill[gray] {(A1) circle (5pt) node {}}
			{(A2) circle (5pt) node {}}
			{(A3) circle (5pt) node {}}
			{(A4) circle (5pt) node {}}
			{(B1) circle (5pt) node {}}
			{(B2) circle (5pt) node {}}
			{(B3) circle (5pt) node {}}
			{(B4) circle (5pt) node {}}
			{(C1) circle (5pt) node {}}
			{(C2) circle (5pt) node {}}
			{(C3) circle (5pt) node {}}
			{(C4) circle (5pt) node {}}
			{(D1) circle (5pt) node {}}
			{(D2) circle (5pt) node {}}
			{(D3) circle (5pt) node {}}
			{(D4) circle (5pt) node {}};
\draw[thick]{(A1) circle (5pt) node {}}
			{(A2) circle (5pt) node {}}
			{(A3) circle (5pt) node {}}
			{(A4) circle (5pt) node {}}
			{(B1) circle (5pt) node {}}
			{(B2) circle (5pt) node {}}
			{(B3) circle (5pt) node {}}
			{(B4) circle (5pt) node {}}
			{(C1) circle (5pt) node {}}
			{(C2) circle (5pt) node {}}
			{(C3) circle (5pt) node {}}
			{(C4) circle (5pt) node {}}
			{(D1) circle (5pt) node {}}
			{(D2) circle (5pt) node {}}
			{(D3) circle (5pt) node {}}
			{(D4) circle (5pt) node {}};
\node[label={$\sigma_1\cdot\sigma_3\cdot\sigma_2$}, below, yshift = -2.5em] at (D_Mid) {};
\node[label={(c)}, below, yshift = -4em] at (D_Mid) {};
%%%%%%%%%%%%%%%%%%%%%%%%%%%%%%%%%%%%%%%%%%%%%%%%%%%%%%%%%%%%%%%%%%%%%%%%%%%%%%%%%%% SIGMA(1) = SIGMA_3, SIGMA(2) = SIGMA_1, SIGMA(3) = SIGMA_1 %%%%%%%%%
\pgfmathsetmacro{\figgap}{\horzgap*\nodegap}
\coordinate (D1) at ($(0,0)+(\figgap,\vertgap)$); 
\coordinate (D2) at ($(D1)+(1/3*\nodegap,0)$);
\coordinate (D3) at ($(D1)+(2/3*\nodegap,0)$);
\coordinate (D_Mid) at ($0.5*(D2)+0.5*(D3)$);
\coordinate (D4) at ($(D1)+(\nodegap,0)$);
\coordinate (C1) at ($(0,1)+(\figgap,\vertgap)$); 
\coordinate (C2) at ($(C1)+(1/3*\nodegap,0)$);
\coordinate (C3) at ($(C1)+(2/3*\nodegap,0)$);
\coordinate (C4) at ($(C1)+(\nodegap,0)$);
\coordinate (B1) at ($(0,2)+(\figgap,\vertgap)$);
\coordinate (B2) at ($(B1)+(1/3*\nodegap,0)$);
\coordinate (B3) at ($(B1)+(2/3*\nodegap,0)$);
\coordinate (B4) at ($(B1)+(\nodegap,0)$);
\coordinate (A1) at ($(0,3)+(\figgap,\vertgap)$);
\coordinate (A2) at ($(A1)+(1/3*\nodegap,0)$);
\coordinate (A3) at ($(A1)+(2/3*\nodegap,0)$);
\coordinate (A4) at ($(A1)+(\nodegap,0)$);

\draw[name path=line 4,draw=white,ultra thick,
		double distance = 3pt,double=blue!50!red] 
		(D1) -- (C2) -- (C3) -- (B4);
\draw[name path=line 3,draw=white,ultra thick,
		double distance = 3pt,double=blue!50!red] 
		(C1) -- (D2) -- (D3) -- (D4);
\draw[name path=line 2,draw=white,ultra thick,
		double distance = 3pt,double=blue!50!red] 
		(B1) -- (B2) -- (A3) -- (A4);
\draw[name path=line 1,draw=white,ultra thick,
		double distance = 3pt,double=red!50!blue] 
		(A1) -- (A2) -- (B3) -- (C4);
\fill[gray] {(A1) circle (5pt) node {}}
			{(A2) circle (5pt) node {}}
			{(A3) circle (5pt) node {}}
			{(A4) circle (5pt) node {}}
			{(B1) circle (5pt) node {}}
			{(B2) circle (5pt) node {}}
			{(B3) circle (5pt) node {}}
			{(B4) circle (5pt) node {}}
			{(C1) circle (5pt) node {}}
			{(C2) circle (5pt) node {}}
			{(C3) circle (5pt) node {}}
			{(C4) circle (5pt) node {}}
			{(D1) circle (5pt) node {}}
			{(D2) circle (5pt) node {}}
			{(D3) circle (5pt) node {}}
			{(D4) circle (5pt) node {}};
\draw[thick]{(A1) circle (5pt) node {}}
			{(A2) circle (5pt) node {}}
			{(A3) circle (5pt) node {}}
			{(A4) circle (5pt) node {}}
			{(B1) circle (5pt) node {}}
			{(B2) circle (5pt) node {}}
			{(B3) circle (5pt) node {}}
			{(B4) circle (5pt) node {}}
			{(C1) circle (5pt) node {}}
			{(C2) circle (5pt) node {}}
			{(C3) circle (5pt) node {}}
			{(C4) circle (5pt) node {}}
			{(D1) circle (5pt) node {}}
			{(D2) circle (5pt) node {}}
			{(D3) circle (5pt) node {}}
			{(D4) circle (5pt) node {}};
\node[label={$\sigma_3\cdot\sigma_1\cdot\sigma_2$}, below, yshift = -2.5em] at (D_Mid) {};
\node[label={(d)}, below, yshift = -4em] at (D_Mid) {};
%%%%%%%%%%%%%%%%%%%%%%%%%%%%%%%%%%%%%%%%%%%%%%%%%%%%%%%%%%%%%%%%%%%%%%%%%
\end{tikzpicture}
\caption{A braid string $\sigma = \sigma_1\cdot\sigma_3\cdot\sigma_2$ taking place in a varying number of braid steps.
}
\label{fig:braid_relation_example}
\end{figure}
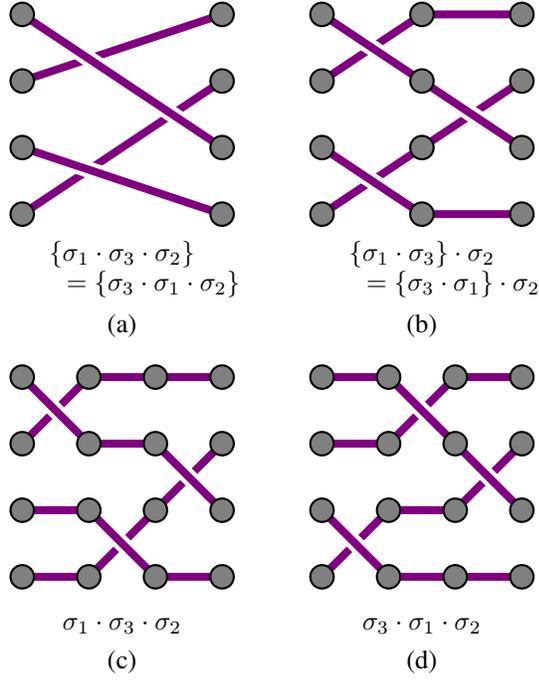

As an example, suppose that a braid string contains the substring of three concatenated generators $\sigma_1\cdot\sigma_3\cdot\sigma_2$. The first two generators, $\sigma_1\cdot\sigma_3$, may take place simultaneously at braid  step 1 without incurring in more than one interaction per agent with another agent, but we would require an additional braid step for $\sigma_2$ in order to avoid multiple interactions in the same step. This is illustrated geometrically in Fig. \ref{fig:braid_relation_example}. In (a), an agent interacts with more than one other agent since all the generators are not at least two indices apart and take place at the same braid step. Note that in the remaining cases (b)-(d), we introduce intermediate braid points while retaining the desired level of interaction, final configuration of the agents, and reducing to pairwise interactions.

Braid strings that satisfy Restriction \ref{res:simult_braids} will have at most interactions involving two agents at any given braid step. So in order to safely execute braids, one need only consider the case when two agents interact. We will devise a strategy for reparameterizing the geometry such that should two agents interact, the resulting parameters are at least $\delta$ from each other.  
By tracking the parameterized paths, the agents' trajectories will be collision-free and braid point feasible. The resulting controllers can be 
combined to satisfy a given braid string.

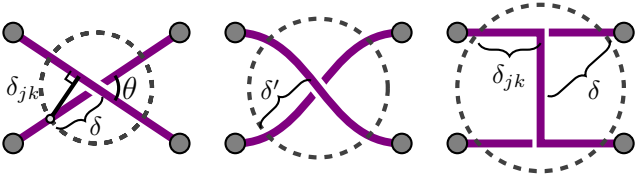
\begin{figure}[t]
\centering
\begin{tikzpicture}[scale=\linewidth/12cm]
%%%%%%%%%%%%%%%%%%%%%%%%%%%%%%%%%%%%%%%%%%%%%%%%%%%%%%%%%%%%%%%%%%%%%%%%%%%%%%%%%%%%%%%%%%%
\draw[name path=line 1,draw=white,ultra thick,double distance = 3pt,double=red!50!blue] (0,0) -- (3,2);
\draw[name path=line 2,draw=white,ultra thick,double distance = 3pt,double=blue!50!red] (0,2) -- (3,0);
\fill[gray] {(0,0) circle (5pt) node {}}
			{(3,0) circle (5pt) node {}}
			{(0,2) circle (5pt) node {}}
			{(3,2) circle (5pt) node {}};
\draw[thick]{(0,0) circle (5pt) node {}}
			{(3,0) circle (5pt) node {}}
			{(0,2) circle (5pt) node {}}
			{(3,2) circle (5pt) node {}};
\draw[name path=Circle1,dashed,ultra thick,name intersections={of=line 1 and line 2,total=\t},darkgray]
    \foreach \s in {1,...,\t}{(intersection-\s) circle (1)};
\coordinate (P) at (intersection-1);
\coordinate (P0) at (0,2);
\draw[thick,name intersections={of=line 1 and Circle1, name=k}] (k-2) coordinate (P2);
\coordinate (P1) at ($(P0)!(P2)!(3,0)$);
\node[left,xshift=-0.66em,yshift=0.15em] (blah) at  ($(P1)!0.4!(P2)$) {$\delta_{jk}$};%\scriptstyle
	\draw[ultra thick] (P1) -- (P2);
	\draw[thick] ($(P1)!5pt!(P2)$)
		--($ (P1)!2!($($(P1)!5pt!(P2)$)!.5!($(P1)!5pt!(P0)$)$) $)
	    --($(P1)!5pt!(P0)$);
\fill[gray!25,name intersections={of=line 1 and Circle1, name=k}] (k-2) circle (2pt);
\draw[thick,name intersections={of=line 1 and Circle1, name=k}] (k-2) circle (2pt);
\draw [thick,decorate,decoration={brace,amplitude=4pt,mirror,raise=4pt},yshift=0pt]
(P2) -- (P) node [black,midway,xshift=0.3cm,yshift=-0.3cm] {$\delta$};
%%%%%%%%%%%%%%%%%%%%%%%%%%%%%%%%%%%%%%%%%%%%%%%%%%%%%%%%%%%%%%%%%%%%%%%%%%%%%%%%%%%%%%%%%%%
\tkzDefPoint(3,2){C}
\tkzDefPoint(1.5,1){B}
\tkzDefPoint(3,0){A}
\tkzMarkAngle[size=0.4 cm,very thick,mkpos=10](A,B,C)
\tkzLabelAngle[pos=0.6](A,B,C){\large $\theta$}
%%%%%%%%%%%%%%%%%%%%%%%%%%%%%%%%%%%%%%%%%%%%%%%%%%%%%%%%%%%%%%%%%%%%%%%%%%%%%%%%%%%%%%%%%%%
\def\DrawingGap{1}
\pgfmathsetmacro{\DrawingGapTwo}{\DrawingGap+4}
\draw[name path=line 3,draw=white,ultra thick,double distance = 3pt,double=red!50!blue] 
				(3+\DrawingGap,0) to[out=0,in=180]  (6+\DrawingGap,2);
\draw[name path=line 4,draw=white,ultra thick,double distance = 3pt,double=blue!50!red] 
				(3+\DrawingGap,2) to[out=0,in=180]  (6+\DrawingGap,0);
\fill[gray] {(3+\DrawingGap,0) circle (5pt) node {}}
			{(6+\DrawingGap,0) circle (5pt) node {}}
			{(3+\DrawingGap,2) circle (5pt) node {}}
			{(6+\DrawingGap,2) circle (5pt) node {}};
\draw[thick]{(3+\DrawingGap,0) circle (5pt) node {}}
			{(6+\DrawingGap,0) circle (5pt) node {}}
			{(3+\DrawingGap,2) circle (5pt) node {}}
			{(6+\DrawingGap,2) circle (5pt) node {}};
\draw[name path=Circle2,dashed,ultra thick,name intersections={of=line 3 and line 4,total=\t},darkgray]
    \foreach \s in {1,...,\t}{(intersection-\s) circle (1.25)};
\coordinate (C2) at (intersection-1);
\path[name intersections={of=line 3 and Circle2, name=m}] (m-2) coordinate (D2);
\draw[thick,decorate,decoration={brace,amplitude=4pt,raise=4pt},yshift=0pt]
(D2) -- (C2) node [black,midway,xshift=-0.3cm,yshift=0.3cm] {$\delta'$};
%%%%%%%%%%%%%%%%%%%%%%%%%%%%%%%%%%%%%%%%%%%%%%%%%%%%%%%%%%%%%%%%%%%%%%%%%%%%%%%%%%%%%%%%%%%
\coordinate (C3) at (4.5+\DrawingGapTwo,1);
\draw[name path=line 5,draw=white,ultra thick,
	  double distance = 3pt,double=red!50!blue] 
				(3+\DrawingGapTwo,0) -| (C3) |- (6+\DrawingGapTwo,2);
\draw[name path=line 6,draw=white,ultra thick,
	  double distance = 3pt,double=blue!50!red] 
				(3+\DrawingGapTwo,2) -| (C3) |- (6+\DrawingGapTwo,0);
\fill[gray] {(3+\DrawingGapTwo,0) circle (5pt) node {}}
			{(6+\DrawingGapTwo,0) circle (5pt) node {}}
			{(3+\DrawingGapTwo,2) circle (5pt) node {}}
			{(6+\DrawingGapTwo,2) circle (5pt) node {}};
\draw[thick]{(3+\DrawingGapTwo,0) circle (5pt) node {}}
			{(6+\DrawingGapTwo,0) circle (5pt) node {}}
			{(3+\DrawingGapTwo,2) circle (5pt) node {}}
			{(6+\DrawingGapTwo,2) circle (5pt) node {}};
\draw[name path=Circle3,dashed,ultra thick,darkgray]
    {(C3) circle (1.5)};
\path[name intersections={of=line 6 and Circle3, name=m2}] (m2-1) coordinate (D3);
\path[name intersections={of=line 5 and Circle3, name=m2}] (m2-2) coordinate (D4);
\draw[thick,decorate,decoration={brace,amplitude=4pt,raise=4pt},yshift=0pt]
(D4) -- (C3) node [black,midway,xshift=0.3cm,yshift=-0.4cm] {$\delta$};
\draw[thick,decorate,decoration={brace,amplitude=5pt,raise=4pt,mirror},yshift=0pt]
(D3) -- ($(C3)+(0,1)$) node [black,midway,xshift=0.0cm,yshift=-0.6cm] {$\delta_{jk}$};
%%%%%%%%%%%%%%%%%%%%%%%%%%%%%%%%%%%%%%%%%%%%%%%%%%%%%%%%%%%%%%%%%%%%%%%%%%%%%%%%%%%%%%%%%%%
\end{tikzpicture}
\caption[Safety separation region for three different-geometry braids]%
{Safety separation region for three different-geometry,
two-agent braids. For straight lines (left) or ``city-block''-like (right), the distance $\delta$ can
be computed exactly. For arbitrary curves (center), the distance
$\delta'$ can be selected conservatively.\label{fig:SafetySeparation}}
\end{figure}

Consider the geometric path agent $j$ must follow at step $i$ given by $\gamma_{i}^{j}:[0,1]\to\R^2$  with $\gamma_{i}^{j}(0) = \xi(i-1,j)$ and $\gamma_{i}^{j}(1) = \xi(i,j)$. Let $\Delta$ be the arclength of this path, given by
\begin{align*}
\Delta = \int_0^1\sqrt{\left(\dot{\gamma}_{i}^{j}(p)\right)^T\dot{\gamma}_{i}^{j}(p)}\,\mathrm{d}p.
\end{align*}
We wish to find a parameterization of the strand geometries such that the parameters of two intersecting strands, thought of as a virtual vehicles that live on their geometries, are collision-free. The strategy to do so will be to impose constraints on the
agents separation from the path intersection at a specified time. To ensure braid-point feasibility, the strategy will be to impose constraints on the time in which the agents must
get from the beginning of one braid step to the end of that step. To that end, we propose the following constrained optimization problem
\begin{align}\label{eq:braid_param_optimization}
(v_j^\star,v_k^\star) &= \argmin_{(v_j,v_k)} J(v_j,v_k) = \argmin_{(v_j,v_k)}\frac{1}{2}\int_{t_{i-1}}^{t_i} v_j^2 + v_k^2 \,\mathrm{d}\tau
\end{align}
{subject to}
\begin{align*}
&\dot{p}_j=v_j, && p_j(t_{i-1})=0, 
\\
&p_j\left(\frac{t_{i-1}+t_i}{2}\right)=\frac{\Delta+\delta}{2\Delta}, && p_j(t_i) = 1,
\intertext{and}
&\dot{p}_k=v_k, && p_k(t_{i-1})=0, 
\\ 
&p_k\left(\frac{t_{i-1}+t_i}{2}\right)=\frac{\Delta-\delta}{2\Delta}, && p_k(t_i) = 1.
\end{align*}
The constraints at $t_{i-1}$ and $t_i$ will ensure that the reparameterization is braid-point feasible. For collision avoidance we will define a safety separation region, as depicted in Fig. \ref{fig:SafetySeparation}, to be the region from the intersection point to the point where it is possible for the two agents to be within $\delta_{jk}$ of each other, i.e., the set $\left[\frac{\Delta-\delta}{2\Delta},\frac{\Delta+\delta}{2\Delta}\right]$ for $\delta\in[0,\Delta]$ such that $\big\|\gamma_i^j(p_j)-\gamma_i^k(p_k)\big\|\geq\delta_{jk}$ for all $p_j, p_k\in\left[0,1\right]$, where we let $\delta$ be the distance from the intersection point to the boundary of the safety separation region along the path. Note that if the geometry of the braid strands are straight lines, the distance
$\delta$ can be easily computed by $\delta= \delta_{jk}\csc(\theta)$, where $\theta$ is the
angle between the two intersecting lines, and this would also imply the parameters get as close as $\delta_{jk}\csc\left(\theta/2\right)$. 
Similarly, when the geometry is ``city-block''-like, we can find $\delta=\delta_{jk}+\frac{h}{2(N-1)}$ and the agent get as close as $2\delta_{jk}$.
For more general geometries, $\delta$ may be conservatively selected.
At $\bar{t}_i:=\frac{t_{i-1}+t_i}{2}$, i.e., half-time along the braid step, we will require one parameter to exit the safety separation region as the other one is about to enter it. 
All these cases are illustrated in Fig. \ref{fig:SafetySeparation}.
This way we guarantee that the two parameters are never inside the region simultaneously and thus their separation will always be of at least $\delta_{jk}$.

The optimality conditions for equation \eqref{eq:braid_param_optimization} are given by 
\begin{align*}
\dot{\lambda}_j &= \dot{\lambda}_k = 0, & \dot{p}_j &= -\lambda_j, & \dot{p}_k &= -\lambda_k, \\
 p_j(\bar{t}_i) &= \tfrac{\Delta+\delta}{2\Delta}, & p_k(\bar{t}_i) &= \tfrac{\Delta-\delta}{2\Delta}, &%&\mathrm{and} &
p_j({t}_i) &= p_k({t}_i) = 1.
%
%\intertext
%{for agent $j$, and similarly for agent $k$}
%
%\dot{\lambda}_k &= 0, & \dot{p}_k &= -\lambda_k, & p_k(\bar{t}_i) &= \frac{\Delta-\delta}{2\Delta}, &%&\mathrm{and} &
%p_k({t}_i) &= 1.
\end{align*}
Since $\dot{\lambda}_j = \dot{\lambda}_k = 0$, we will get that  ${\lambda}_j$ and ${\lambda}_k$ are piecewise constant. By using the midway condition, we have that for $t\in(t_{i-1},\bar{t}_i]$
\begin{multline*}
p_j(\bar{t}_i)=\int_{t_{i-1}}^{\bar{t}_i}v_j\,\mathrm{d}\tau = \frac{\Delta+\delta}{2\Delta}
\\
\Rightarrow   \qquad v_j({\bar{t}_i}-{t_{i-1}})= \frac{\Delta+\delta}{2\Delta}
\\
\Rightarrow  \qquad v_j = \frac{1}{\Delta}\frac{\Delta+\delta}{({{t}_i}-{t_{i-1}})}
\end{multline*}
and similarly for $t\in(\bar{t}_{i},t_i]$, the terminal condition tells us that
\begin{multline*}
p_j({t}_i)=\int_{\bar{t}_i}^{t_{i}}v_j\,\mathrm{d}\tau+\frac{\Delta+\delta}{2\Delta} = \frac{\Delta}{\Delta}
\\
\Rightarrow \qquad v_j({{t}_i}-\bar{t}_i) = \frac{1}{2\Delta}\left({2\Delta}-{\Delta+\delta}\right)
\\
\Rightarrow \qquad v_j = \frac{1}{\Delta}\frac{\Delta-\delta}{({{t}_i}-{t_{i-1}})}
\end{multline*}
Note that for agent $k$, the signs are reversed on the numerator. As such, the resulting braid parameterization will have velocities given by
\begin{align}\label{eq:braid_parameterization_dynamics}
\dot{p}_j(t) = \left\{\begin{array}{ll}
\frac{1}{\Delta}\frac{\Delta\pm\delta}{({{t}_i}-{t_{i-1}})}& t\in\left(t_{i-1},\bar{t}_i\right]
\\
\frac{1}{\Delta}\frac{\Delta\mp\delta}{({{t}_i}-{t_{i-1}})}& t\in\left(\bar{t}_i,t_i\right]
\end{array}\right.
\end{align}
where the sign in the numerator is determined by the interpretation given to the braid strand going ``under'' (e.g., the agent crosses the intersection point first) or ``over'' (e.g., the agent crosses the intersection point second), and in cases where there are no intersections we set $\delta=0$ in the numerator.

%%%%%%%%%%%%%%%%%%%%%%%%%%%%%%%%%%%%%%%%%%%%%%%%%%%%%%%%%%%%%%%%%%%%%%%%%%%%%%%%%%%%%%%%%%%%
It is possible to come up with an upper bound on the length of the braid attainable under this mixing scheme. Under Restriction \ref{res:simult_braids}, % and the \emph{braid-controller} described in Section \ref{braidControl}
the following theorem provides an upper bound on the mixing limit.

\begin{theorem} \label{thm:MixingLimit}
Given the maximum safety separation $\bar{\delta}$ 
%that defines the distance along the braid from the intersection point to the safety separation region, 
and bounds on the agents' velocities such that $v_j(t) \in[0, v_\mathrm{max}]$ $\forall t,j$, the mixing limit $M^\star$ for $N$-agent braids that can be performed in a space of height $h$ and length $\ell$ in time $T$ is bounded above by
%\begin{equation*}
% M \leq \frac{(N-1)(v_{max}T-\ell)}{((N-1)\delta+h)}
%\end{equation*}
\begin{multline*}
M^\star\leq\min\left\{\frac{\ell\sqrt{4h^2-\bar\delta^2\left(N-1\right)^2}}{\bar\delta h},\right.\\
\left.\frac{\left({N-1}\right)\left(v_\mathrm{max}T-\left(\ell+\bar\delta\right)\right)}{h}-\frac{1}{2}\right\}.
\end{multline*}
\end{theorem}

\noindent
{\it Proof:}\\
Consider $\sigma\in\Sigma_N^M$. Since at any step an agent can either interact with another agent or move straight ahead, the maximum mixing will be achieved if an agent interacts with another agent at every step. Thus, at every step it must be enforced that there are no collisions. Assuming %a uniform distribution of the braid points and 
a uniform partition of the time window, the arclength-normalized parameter velocity will be given by
\begin{align*}
v_j=\left\{\begin{array}{ll}
	\left(\frac{M\Delta\pm M\delta}{T\Delta}\right) & \mbox{if } t \in \left(\frac{i-1}{M}T,\frac{2i-1}{2M}T\right] \\
	\left(\frac{M\Delta\mp M\delta}{T\Delta}\right) & \mbox{if } t \in \left(\frac{2i-1}{2M}T,\frac{i}{M}T\right]
\end{array}\right.
\end{align*}
for $i = 1,2,\ldots,M$, where the sign on the numerator depends on the interpretation of whether the ``strand" goes over or under, $\Delta$ is the arclength of the strand geometry connecting two braid point, assumed equal at each braid step and for both agents due to symmetry, and $\delta$ is the safety separation distance along the braid as described above. The total braid path length $\Delta$ heavily depends on the geometry of path. However, for the sake of obtaining bounds on the mixing limit, we may assume that the braid points are uniformly distributed in height and length, such that the length of 
sufficiently regular paths will be bounded by
\begin{equation}\label{eq:bounds_arclength}
\sqrt{\left(\frac{h}{N-1}\right)^2+\left(\frac{\ell}{M}\right)^2}\leq\Delta \leq \frac{h}{N-1}+\frac{\ell}{M}
\end{equation}
where the lower 
bound assumes straight lines connecting the braid points and the upper bound assumes ``city block''-like paths stepping midway between the two points.
But after normalizing the bounds on the parameter velocity, we see that
\begin{equation}\label{eq:bounds_velocity}
0\leq\frac{M\left({\Delta}-{\delta}\right)}{T{\Delta}}\leq\frac{M\left({\Delta}+{\delta}\right)}{T{\Delta}}\leq \frac{v_\mathrm{max}}{\Delta}.
\end{equation}
The lower bound of \eqref{eq:bounds_velocity} tells us that for the parameter to not go backwards we need 
%\begin{align}\label{eq:safety_separation_arclength_ratio}
${\delta}\leq{\Delta}$. To ensure this, we set $\delta=\bar\delta\csc\theta\leq\sqrt{\left(\tfrac{h}{N-1}\right)^2+\left(\tfrac{\ell}{M}\right)^2}\leq\Delta$, and since we require $\bar\delta\leq \tfrac{h}{N-1}$ for collision-free braid points, using the geometric relationships
to solve for $M$ yields
\begin{align*}
M\leq\frac{\ell\sqrt{4h^2-\bar\delta^2\left(N-1\right)^2}}{\bar\delta h}.
\end{align*}
Similarly,
the right-hand side inequality of \eqref{eq:bounds_velocity} tells us that $\Delta\leq\tfrac{v_\mathrm{max}T-\delta}{M}=\tfrac{v_\mathrm{max}T-\left(\frac{1}{2}\left(\frac{h}{N-1}\right)+\bar\delta\right)}{M}$, and to ensure this we set $\Delta\leq\frac{h}{N-1}+\frac{\ell}{M}\leq\tfrac{v_\mathrm{max}T-\left(\frac{1}{2}\left(\frac{h}{N-1}\right)+\bar\delta\right)}{M}$. 
Solving for $M$  yields
\begin{equation*}
 M \leq \frac{\left({N-1}\right)\left(v_\mathrm{max}T-\left(\ell+\bar\delta\right)\right)}{h}-\frac{1}{2}.
\end{equation*}
Thus
\begin{multline*}
M^\star\leq\min\left\{\frac{\ell\sqrt{4h^2-\bar\delta^2\left(N-1\right)^2}}{\bar\delta h},\right.\\
\left.\frac{\left({N-1}\right)\left(v_\mathrm{max}T-\left(\ell+\bar\delta\right)\right)}{h}-\frac{1}{2}\right\}.\tag*{$\blacksquare$}
\end{multline*}

\begin{figure}[t]
	\begin{center}
	\begin{tikzpicture}[scale=\linewidth/10cm]
		\def\ell{2}
		\def\h{4}
		\def\d{0.13}
		\def\vmax{2}
		\def\T{30}
		\begin{axis}[scale=0.9, xlabel={Number of Agents $N$}, 
					 compat=newest,
					 ylabel={Time Alloted $T$ (s)},
					 zlabel={Mixing Limit Bound $M$}]
			\addplot3[surf,samples=50,domain=2:29, y domain=1:60] 
			{floor(min(\ell*sqrt(4*pow(\h,2)-pow(\d,2)*pow(x-1,2))/\d/\h,
			(x-1)*(\vmax*y-\ell-\d)/\h-1/2))};
		\end{axis}
		\end{tikzpicture}
	\end{center}
	\caption{Upper bound on the \emph{Mixing Limit} presented in Theorem \ref{thm:MixingLimit} for parameters $\ell=2~m$, $h=4~m$, $\delta=0.13~m$, $v_\mathrm{max}=2~m/s$}\label{fig:MixingLimitThm}
\end{figure}
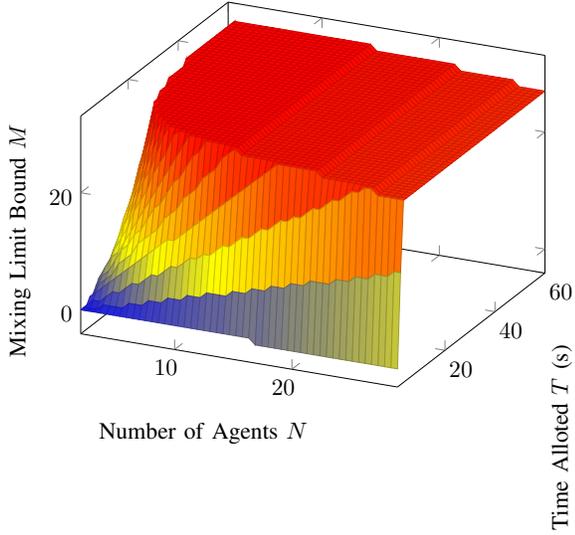

Theorem \ref{thm:MixingLimit} provides a compact expression to obtain an upper bound on the mixing limit that abstracts away strand geometry. It provides a notion of the whether or not desirable mixing levels are achievable in the space, regardless of what the actual movement patterns to achieve these mixing levels are (encoded in the braid string of length $M\leq M^\star$). Fig. \ref{fig:MixingLimitThm} includes a plot of the bound on the mixing limit for varying number of agents and time window size.

%%%%%%%%%%%%%%%%%%%%%%%%%%%%%%%%%%%%%%%%%%%%%%%%%%%%%%%%%%%%%%%%%%%%%%%%%%%%%%%

\subsection{Non-Rectangular Regions}
\label{sec:ArbRegions}
\begin{figure}[!t]%
\centering
\subfloat[Three agents braiding on a curved region.]{\centering
\includegraphics[width=0.75\columnwidth]{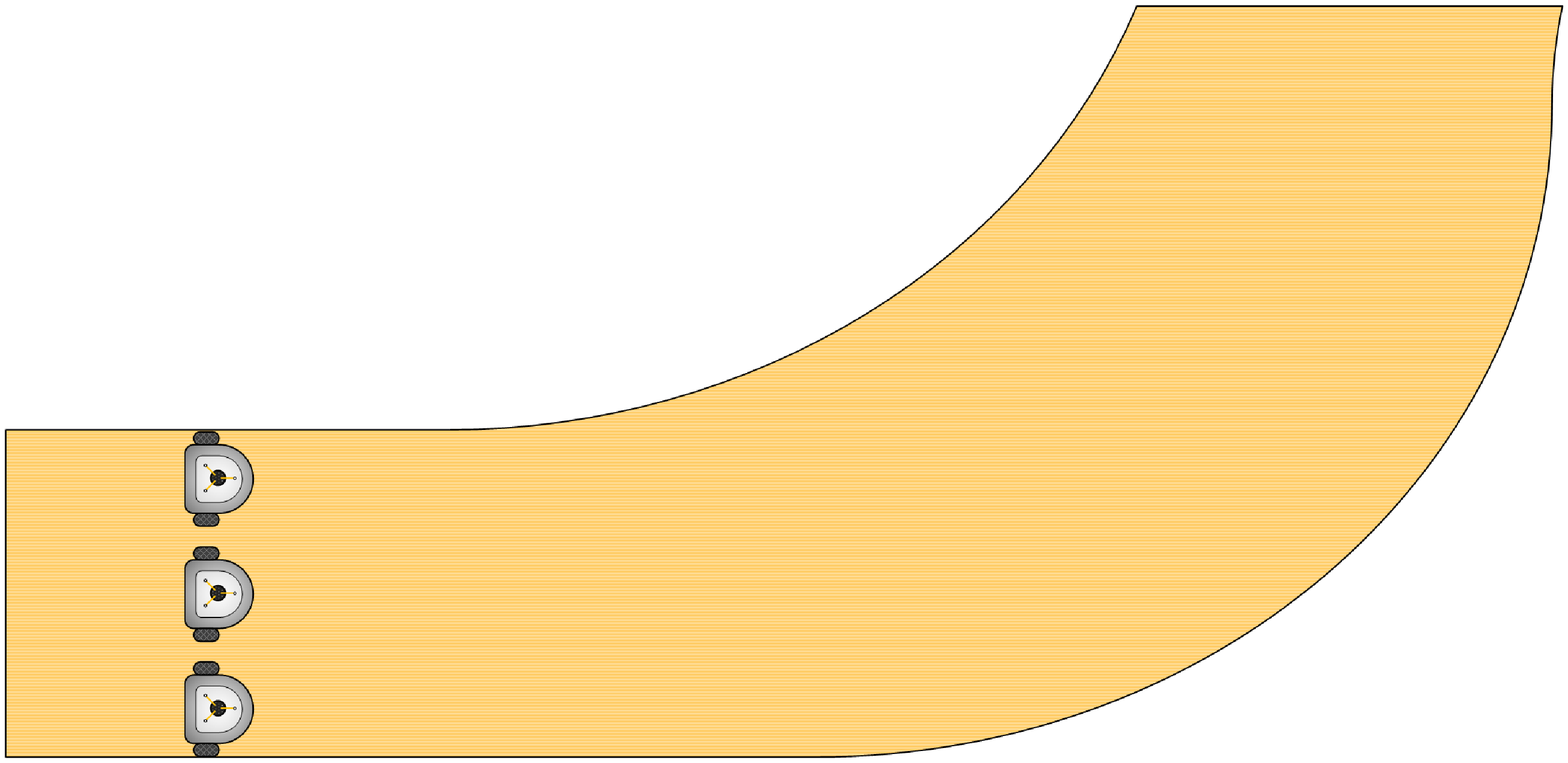}%
}
\\
\centering
\hfill
\subfloat[Two agents need to move from the left-most circles to the right-most circles along the curved region.\label{fig:QuadBraid}]{\centering
\includegraphics[width=0.375\columnwidth]{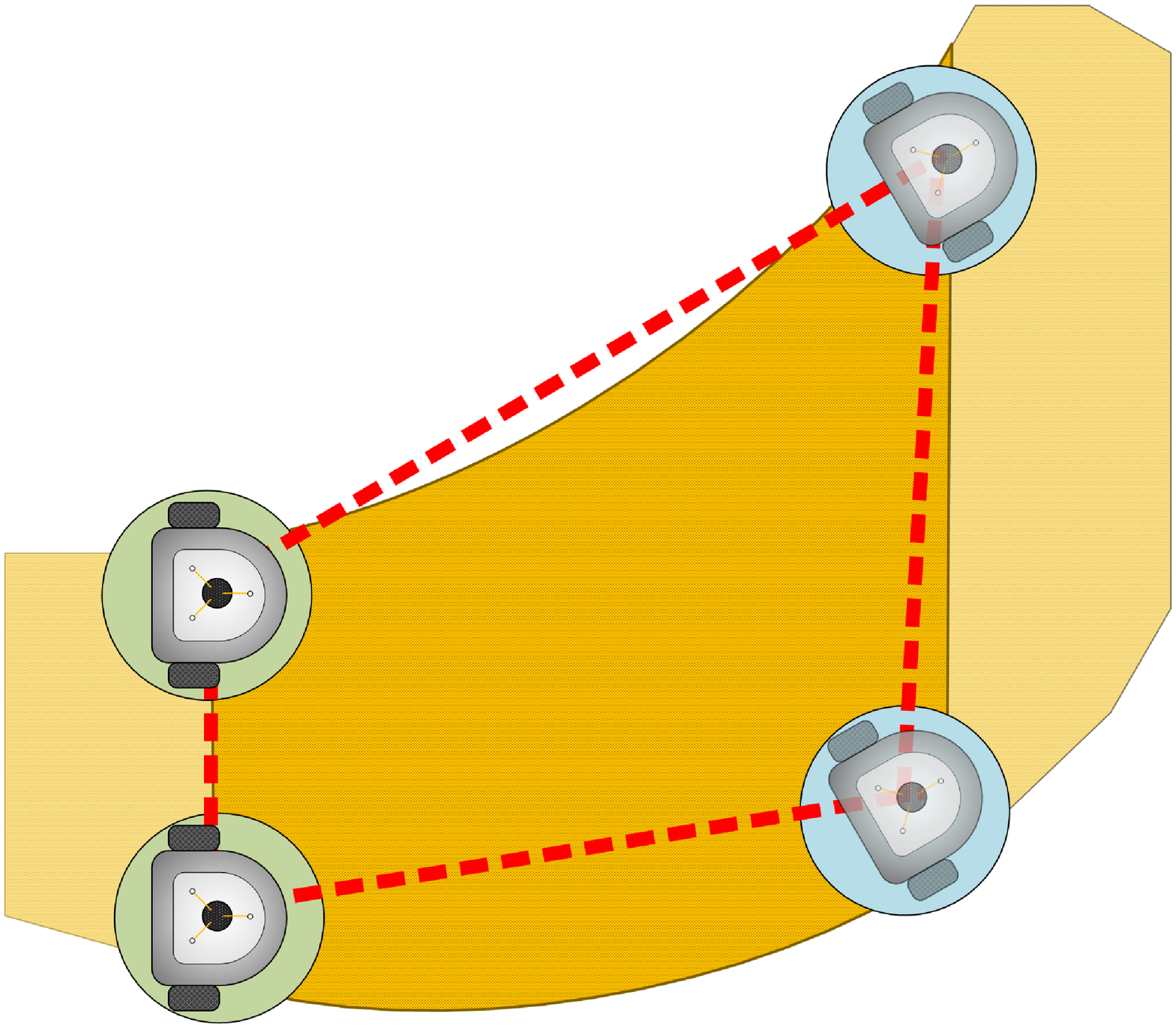}%
}\hfill
\subfloat[The longer the braid length in this curved road segment, the closer the quadrilaterals resemble the curved region.\label{fig:QuadBraid2}]{\centering
\includegraphics[width=0.375\columnwidth]{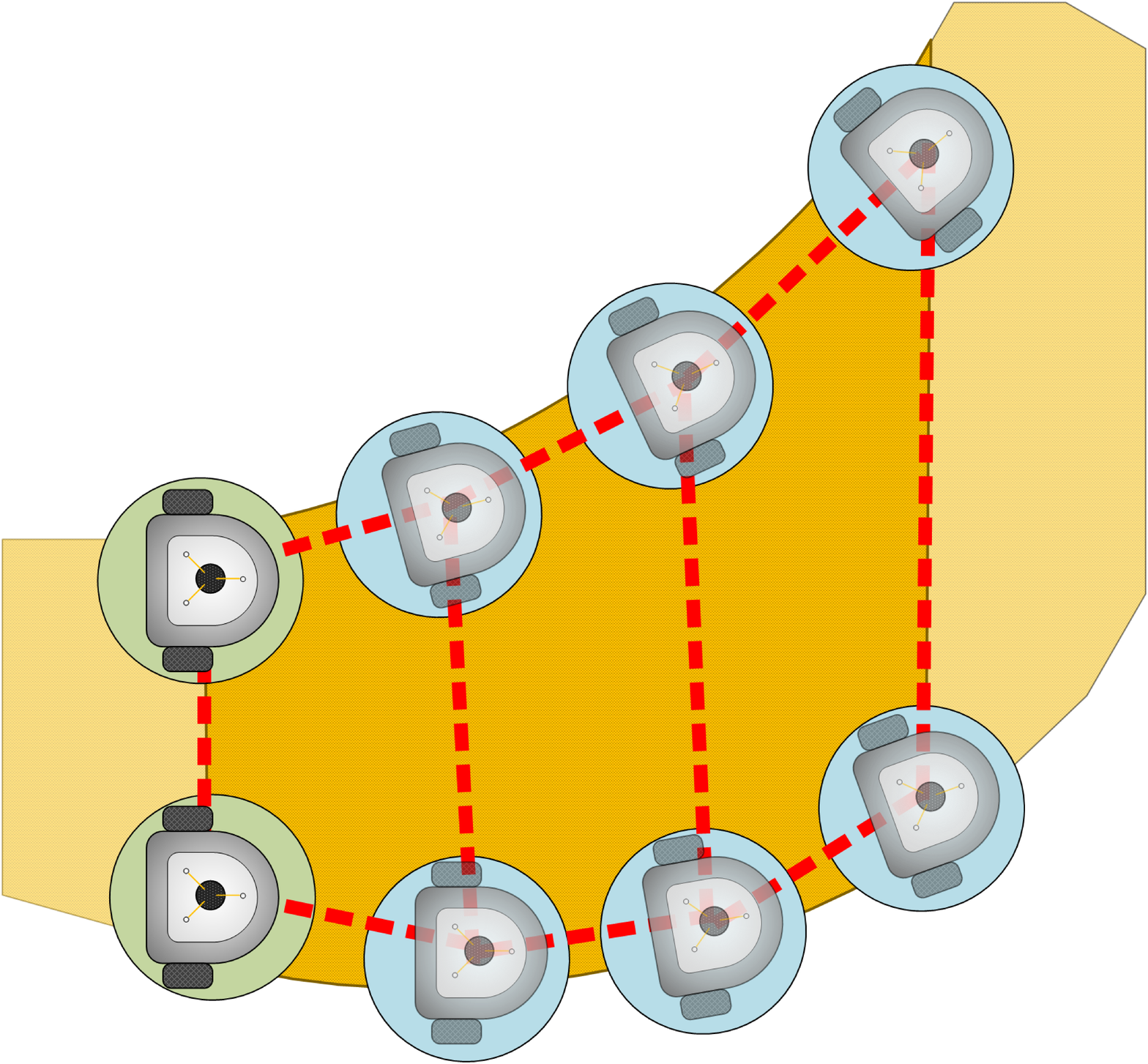}
}
\hfill\ 
\caption{Agents braiding on a region that curves.}%
\label{fig:ArbPath}
\end{figure}

Up to this point, only the problem of braiding on a rectangular region of height $h$ and length $l$ has been considered. On this region, the braid points were uniformly distributed along both dimensions and bounds on the mixing limit were provided through the use of the proposed braid controller. In this section, the scheme is extended to more generally shaped regions, e.g., the road on Fig. \ref{fig:ArbPath}. As has been done previously, discussion begins by first considering the two agent case.

Consider the two agents attempting to perform a braid of length one on the arbitrarily curved region on Fig. \ref{fig:QuadBraid}.
Connecting the braid points together results in the quadrilateral depicted in the red dotted line. Let this quadrilateral be considered as the space where the agent needs to perform a braid of length one, rather than the curved region itself.
Note that as longer length braids are included in this road segment, other quadrilaterals appended together will be obtained which approximate the road slightly better. Since it is of interest to obtain mixing strategies near the mixing limit, as longer length braids are included in this road segment, better approximations of the curved region will be obtained by these composition of quadrilaterals. This is depicted in Fig. \ref{fig:QuadBraid2}.

\begin{figure}[t]\centering
\subfloat[The curved region will be mapped to the rectangular region of known width and height where control design will take place.\label{fig:StraightenedPath}]{\centering
\includegraphics[width=0.9\columnwidth]{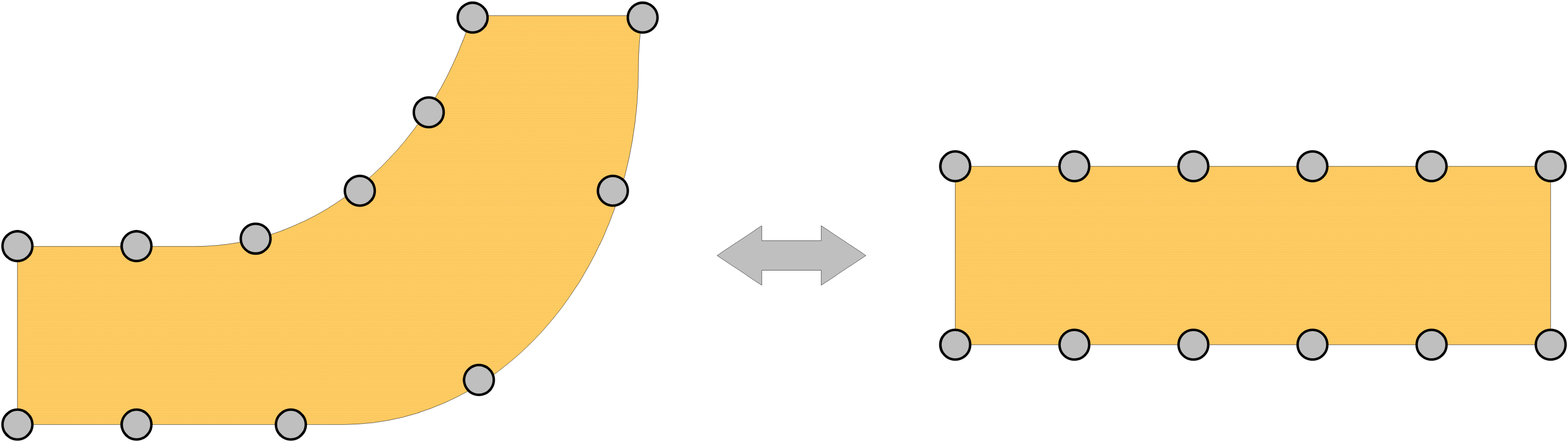}
}\\
\subfloat[The bijective transformation $\mathcal{T}$ maps points in the rectangle to points in the quadrilateral. Both shapes are defined by the braid points which determine the corners.\label{fig:Quad2Rect}]{\centering\parbox{0.9\linewidth}{\centering
\includegraphics[width=0.7\columnwidth]{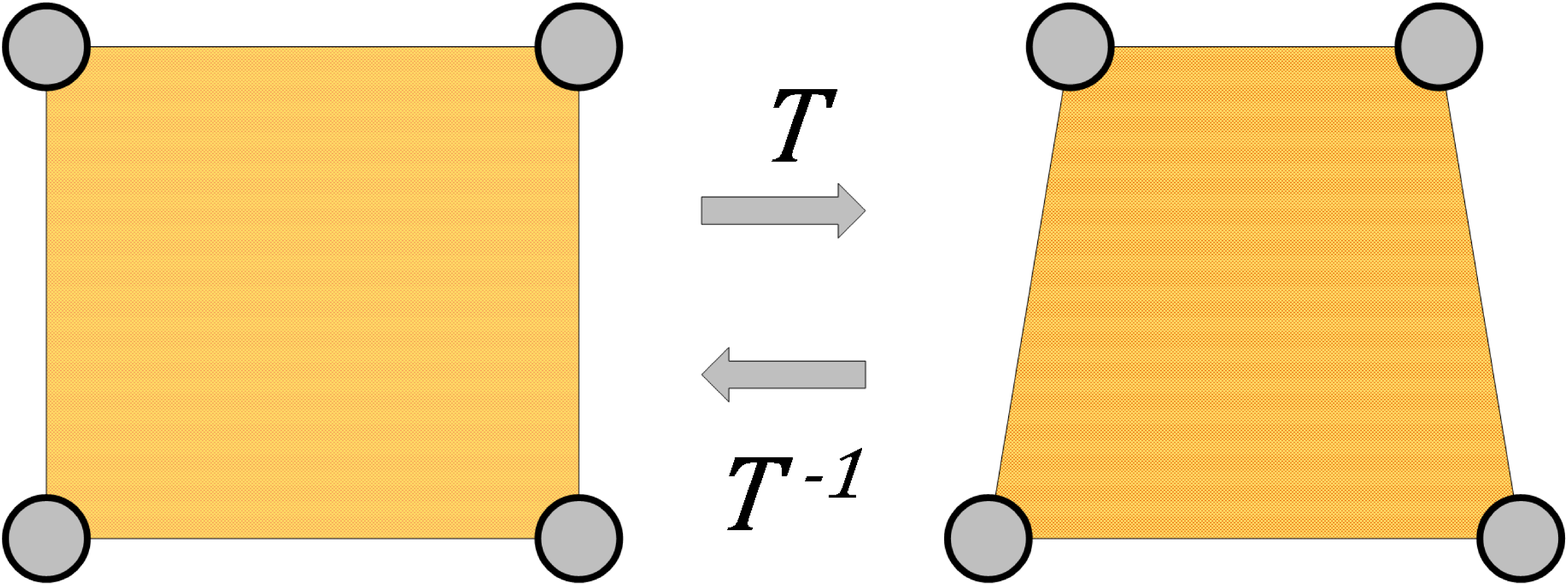}}
}
\caption{Rectangular and non-rectangular regions.}
\end{figure}

The strategy for performing a mixing strategy in curved region will be to transform the curved region into a straightened rectangular region of known height and length, as illustrated in Fig. \ref{fig:StraightenedPath}. In this way the braid controller can be fashioned as in previous sections and the resulting braid controller can be transformed back into the actual curved region.
After distributing the braid points on both the curved region and the rectangular region, as in Fig.  \ref{fig:StraightenedPath}, the next step is to find a transformation to map between these two regions.

Let agents $j$ and $k$ interact in at braid step $i$. Denote ${\cal{S}}^q_{i,j}$ to be the quadrilateral formed by connecting together the braid points $\xi(i-1,j), \xi(i-1,k), \xi(i,j)$, and $\xi(i,k)$, like the one depicted in Fig. \ref{fig:QuadBraid}. Let ${\cal{S}}^r_{i,j}$ be a rectangular plane of specified height and length whose corners are given by $\xi_r(i-1,j), \xi_r(i-1,k), \xi_r(i,j)$, and $\xi_r(i,k)$.

With knowledge of these braid points and through the use of a projective transform as in \cite{criminisi1999plane}, it is possible to obtain a local diffeomorphism that maps from a rectangle of specified height and length to the convex arbitrarily shaped quadrilateral, i.e.,
$\mathcal{T}_{i,j}:{\cal{S}}^r_{i,j}\to{\cal{S}}^q_{i,j}$.

\begin{figure*}[t]
\centering
%\begin{subfigure}{0.49\textwidth}\centering
\subfloat[Simulation Time $5.04$ seconds.]{\centering
\includegraphics[width=0.45\textwidth, clip=true,trim=25 40 30 110]{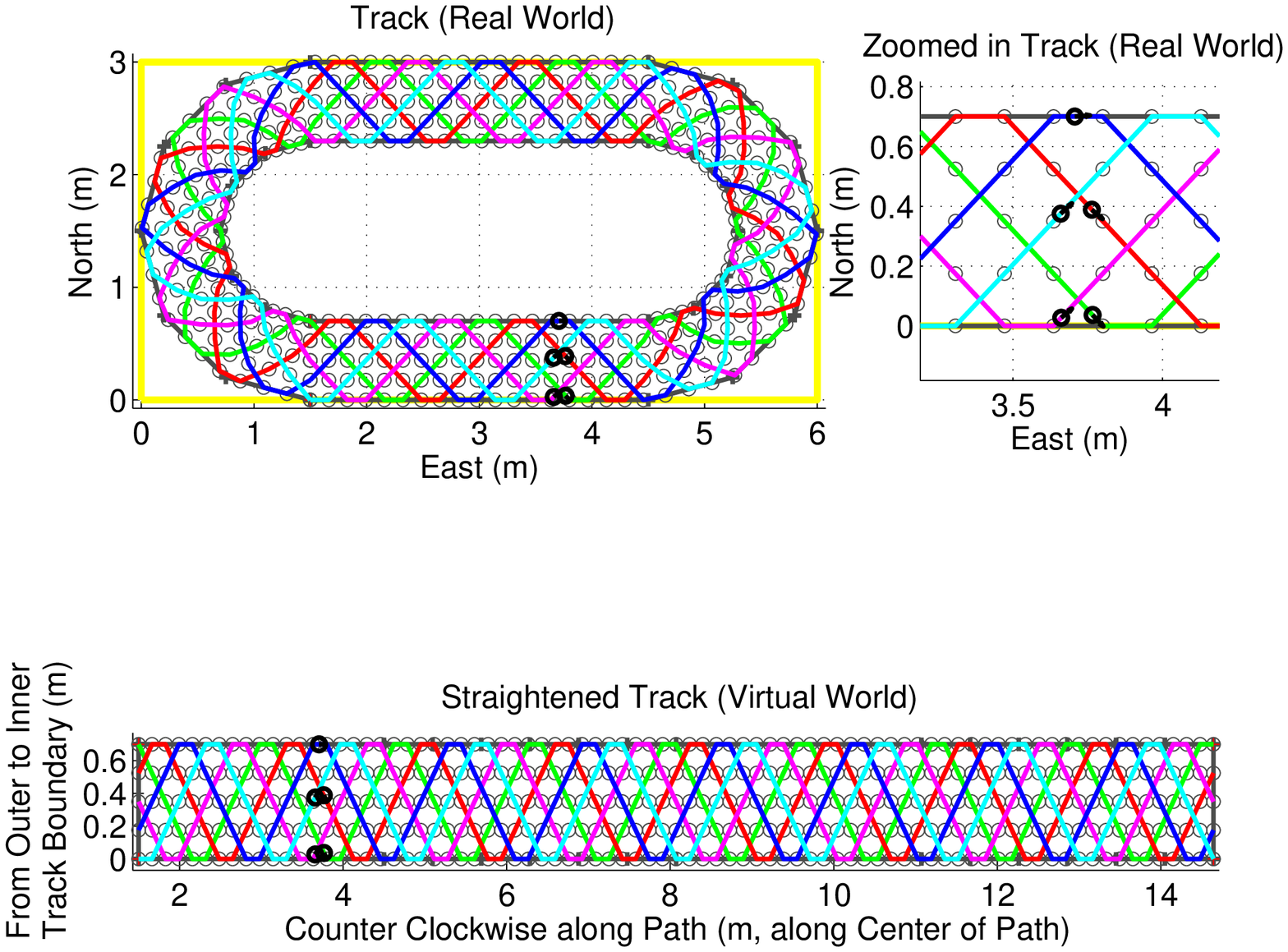}
}\hfill
\subfloat[Simulation Time $9.57$ seconds.]{\centering
\includegraphics[width=0.45\textwidth, clip=true,trim=25 40 30 110]{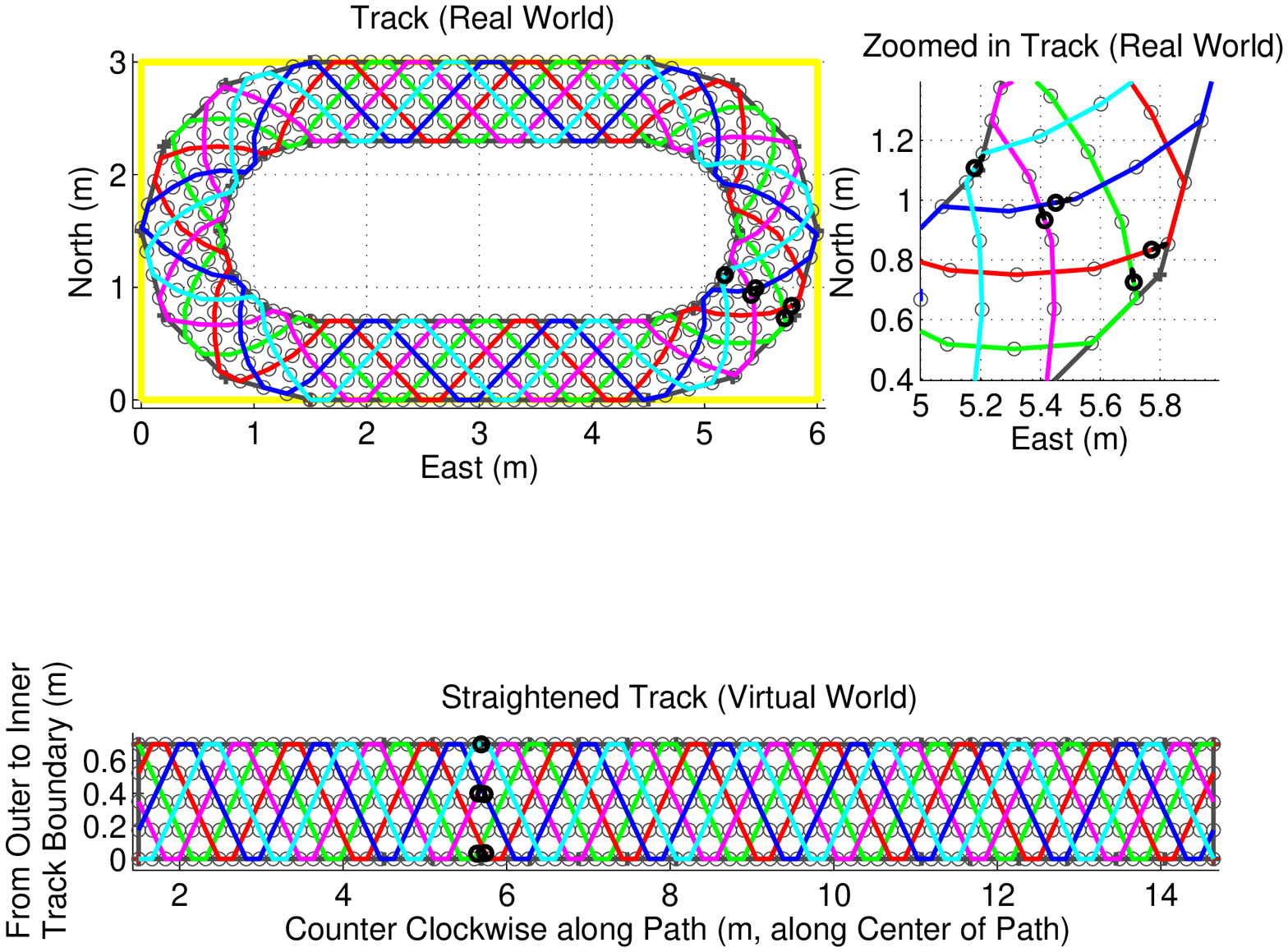}
}
\caption{Five agents performing a mixing strategy given by a braid of length 80. The top left plot represents the actual curved region the agents are mixing in and the top right is a close-up of the agents. The lower plot is the virtual ``straightened'' rectangular region where the design of the braid took place.}
\label{fig:BraidingTrack}
\end{figure*}

Note that by selecting transforms based on the corners of these quadrilaterals, a continuous curve that spans across the boundary between ${\cal{S}}^r_{i,j}$ and ${\cal{S}}^r_{i+1,j}$
might be mapped to a discontinuous curve that spans across the boundary between ${\cal{S}}^q_{i,j}$ and ${\cal{S}}^q_{i+1,j}$  when transformed using $\mathcal{T}_{i,j}$ and $\mathcal{T}_{i+1,j}$ in their respective spaces. However, there will certainly be continuity in the mapping of curves passing through the braid points, since these points are shared by the quadrilaterals and are used to compute the transforms, i.e.,
$\mathcal{T}_{i,j}\left(\xi_r(i,j)\right)=\xi(i,j)=\mathcal{T}_{i+1,j}\left(\xi_r(i,j)\right)$.

Recall the braid controller proposed for the rectangular region. For agent $j$, at braid step $i$, this was given by
\begin{align*}
v_j(t) = \left\{ \begin{array}{ll}
         \frac{1}{\Delta}\left(\frac{\Delta-\delta}{t_i-t_{i-1}}\right) & \mathrm{if}~ t\in \left(t_{i-1},\bar{t}_i\right],\\
         \frac{1}{\Delta}\left(\frac{\Delta+\delta}{t_i-t_{i-1}}\right) & \mathrm{if}~ t\in \left(\bar{t}_i,t_i\right].\end{array} \right.
\end{align*}
where it is expected that $\mathcal{T}^{-1}_{i,j}\left(x\left(t_{i-1}\right)\right)=\xi_r(i-1,j)$ and $\mathcal{T}^{-1}_{i,j}\left(x\left(t_{i}\right)\right)=\xi_r(i,j)$. In the expression, $\Delta$ corresponds to the length of the geometric path agent $j$ must follow to move between $\xi_r(i-1,j)$ and $\xi_r(i,j)$, while $\delta$ corresponds to the distance along the path 
agent $j$ 
switches velocities in order to avoid collisions. Note that for a given parameterization of the geometric path $\gamma_{i,r}^j(p)$ in ${\cal{S}}^r_{i,j}$ with parameter $p\in[0,1]$,
the arclength $\Delta$ may be computed as follows
\begin{align*}
\Delta = \int_0^1 \sqrt{\dot{\gamma}_{i,r}^{j\top}(p)\dot{\gamma}_{i,r}^j(p)}\,\mathrm{d}p.
%=\int_a^b \sqrt{\dot{\gamma}_r^{\top}(p)\mathcal{M}\left(\gamma(p)\right)\dot{\gamma}_r(p)}\,\mathrm{d}p
\end{align*}
 If the geometric curve is directly given in ${\cal{S}}^q_{i,j}$ as $\gamma_{i}^j(p)$ such that the curve in ${\cal{S}}^r_{i,j}$ may be parameterized as $\gamma_{i,r}^j(p) = \mathcal{T}_{i,j}^{-1}(\gamma_{i}^j(p))$, then the arclength may be computed by
\begin{align*}
\Delta = \int_0^1 \sqrt{\left(\dot{\gamma}_{i}^{j}(p)\right)^{\top}\mathcal{M}(\gamma_{i}^j(p))\dot{\gamma}_{i}^j(p)}\,\mathrm{d}p.
\end{align*}
where $\mathcal{M}(\gamma_i^j(p)) = {\left(D\mathcal{T}_{i,j}^{-1}\left(\gamma_i^j(p)\right)\right)}^{\top}D\mathcal{T}_{i,j}^{-1}\left(\gamma_i^j(p)\right)$ and $D\mathcal{T}_{i,j}^{-1}$ is the Jacobian of $\mathcal{T}_{i,j}^{-1}$. In the special case where the geometry is given by straight lines, then
$\Delta =
\norm{\xi_r(i,j)-\xi_r(i-1,j)}$.

Finally, it is of interest to find $\delta$ in order to avoid collisions. If the 
parameterization $\gamma_i^j(p)$ of the curve in ${\cal{S}}^q_{i,j}$ is known, 
then the safety separation ball may be set in ${\cal{S}}^q_{i,j}$ as before and 
the safety separation distance may be computed as
\begin{align*}
\delta = \int_a^b \sqrt{{\left(\dot{\gamma}_i^j(p)\right)}^{\top}\mathcal{M}(\gamma_i^j(p))\dot{\gamma}_i^j(p)}\,\mathrm{d}p.
\end{align*}
where $[a,b]\subset[0,1]$, and if the agent is meant to braid ``over'' (resp. 
``under'') then $\gamma_i^j(a)$ corresponds to the point along the curve where the 
agent enters the safety separation region (resp. where the two curves intersect) 
and $\gamma_i^j(b)$ corresponds to the point along the curve where the two curves 
intersect (resp. where the agent exits the safety separation region). 

In the special case where the geometry is given by straight lines, then by setting
\begin{align*}
\gamma_i^n(p) = (1-p)\xi(i-1,n)+p\xi(i,n),~ p\in[0,1],~ n = j, k.
\end{align*}
as the path agent $j$ and $k$ must follow, 
the intersection point $s$ may be found by setting $s = \gamma_i^j(\pi_j)=\gamma_i^k(\pi_k)$ where 
\begin{align*}
\begin{bmatrix}
\pi_j\\ \pi_k
\end{bmatrix}=
A^{-1}\left(\xi(i-1,k)-\xi(i-1,j)\right)
\end{align*}
with 
$A=
\left[\left(\xi(i,j)-\xi(i-1,j)\right),
-\left(\xi(i,k)-\xi(i-1,k)\right)\right]$.
Recall that the distance in ${\cal{S}}^q_{i,j}$ from the intersection point $s$, for the special case of the geometry being straight lines, was given by $\delta=\delta_{jk}\csc(\theta)$ with $\theta$ being the angle between these two lines, i.e., $\theta = \cos^{-1}\left(\hat{x}_j^{\top}\hat{x}_k\right)$ where $\hat{x}_n$ is the unit vector pointing towards the next point, i.e.,
\begin{align*}
\hat{x}_n&=\frac{\left(\xi(i,n)-\xi(i-1,n)\right)}{\norm{\xi(i,n)-\xi(i-1,n)}},\qquad n = j, k.
\end{align*}
By setting $\hat{\gamma}(p) = (1-p)s\pm p\delta\hat{x}_j$, where sign depends 
on whether the braid goes ``over'' or ``under,'' it is possible to determine 
$\delta$ in the non-rectangular plane directly as
\begin{align*}
\delta_r = \int_0^1\sqrt{\left(\pm \delta\hat{x}_j-s\right)^{\top}\mathcal{M}\left(\hat{\gamma}(p)\right)\left(\pm \delta\hat{x}_j-s\right)}\,\mathrm{d}p.
\end{align*}
With this information, $v_j$ can be found in ${\cal{S}}^r_{i,j}$ for interactions between agents $j$ and $k$. Note that a reparameterization of the path can now be set equivalent to the desired trajectory of agent $j$ by setting $\gamma_{i,d}^j(t) =\gamma_{i,r}^j(p_j(t))$ (see \eqref{eq:braid_parameterization} below). Thus, the braid controller parameter velocity $v_j^q(t)$ for agent $j$ in ${\cal{S}}^q_{i,j}$ will be given by
$v_j^q(t) = \norm{D\mathcal{T}_{i,j}(\gamma_{i,d}^j(t))\dot{\gamma}_{i,d}^j(t)}$.

This strategy was implemented in simulation over the curved region illustrated in Fig. \ref{fig:BraidingTrack}. In the figure, agents are performing the mixing strategy given to them by a braid of length 80 on the curved region (top) and simultaneously on the straightened rectangular region (bottom). The parameters used for this simulation were $\delta_{jk} = 7.7~cm$ and $v_{max} = {1.5~m/s}$ $\forall j,k$, and $T = 30~s$.

%%%%%%%%%%%%%%%%%%%%%%%%%%%%%%%%%%%%%%%%%%%%%%%%%%%%%%%%%%%%%%%%%%%%%%%%%%%%%%%

\section{Implementing Braids}\label{sec:implementing_braids}
Section \ref{sec:braid} approached the problem of multi-robot mixing from an execution level, where given a specification it is possible to synthesize a braid controller that satisfies the specification. In this section we consider the implementation level, that is, we address how to find controllers that are implementable on actual robotic systems that follow the braid controllers found in the previous section.  We then validate the framework by implementing a mixing strategy on a team of six robots.

%%%%%%%%%%%%%%%%%%%%%%%%%%%%%%%%%%%%%%%%%%%%%%%%%%%%%%%%%%%%%%%%%%%%%%%%%%%%%%%%%%%%%%%%%%%%
\subsection{Optimal Tracking Controller}\label{sec:tracking_controller}
We will now utilize the braid parameter velocity to find the braid controller for the agents.
By integrating \eqref{eq:braid_parameterization_dynamics} we obtain the \emph{braid parameterization} of the path $\gamma_i^j$
\begin{align}\label{eq:braid_parameterization}
{p}_j(t) = \left\{\begin{array}{ll}
\frac{t-t_{i-1}}{({{t}_i}-{t_{i-1}})}\frac{\Delta\pm\delta}{\Delta}
& t\in\left(t_{i-1},\bar{t}_i\right]
\\
\frac{t-\bar{t}_i}{({{t}_i}-{t_{i-1}})}\frac{\Delta\mp\delta}{\Delta}+\frac{\Delta\pm\delta}{2\Delta}
& t\in\left(\bar{t}_i,t_i\right].
\end{array}\right.
\end{align}
Assume that agents have single integrator dynamics, i.e., $\dot{x}_j=u_j$ with $y_j=x_j\in\R^2$. The agent's controller will be found by optimally tracking the reparameterized path $\gamma_{i}^{j}(p_j(t))$ to minimize the cost
\begin{align}\label{eq:Tracking_Cost}
J(u_j) = \frac{1}{2}\int_{t_{i-1}}^{t_i} \left(x_j-\gamma_{i}^{j}\right)^TQ \left(x_j-\gamma_{i}^{j}\right)+u_j^TRu_j\,\mathrm{d}\tau 
\end{align}
for $Q=Q^T\succ0$ and $R=R^T\succ0$, with constraints
$\dot{x}_j = u_j$, $x_j(t_{i-1}) = \gamma_{i}^{j}(0)$, and $x_j(t_{i}) = \gamma_{i}^{j}(1)$.
Using the standard variational argument together with Pontryagin's minimum principle, 
the
first order necessary conditions for optimality 
tell us that the optimal tracking controller $u^*_j$ is given by
\begin{align*}
u^*_j &= -R^{-1}\lambda_j
\intertext{where $\lambda$ is the so-called costate and satisfies}
\dot{\lambda}_j &= -Q\left({x}_j-\gamma_i^j\right)
\end{align*}
with unknown terminal condition $\lambda_j(t_i)$.
Suppose that similarly to \cite[Chapter~5.3]{bryson&ho} we can construct $\lambda_j$ as an affine combination of the unknown $\lambda_j(t_i)$ and the state, i.e.,
\begin{align*}
\lambda_j(t) = H(t){x}_j(t) + K(t) \lambda_j(t_i) + E(t)
\end{align*}
and similarly, the terminal state as
\begin{align*}
{x}_j(t_i) &= \xi(i,j) = F(t) {x}_j(t) + G(t) \lambda_j (t_i) + D(t)%\\
%&=F(t_{i-1}) \xi(i-1,j) + G(t_{i-1}) \lambda_j (t_i) + D(t_{i-1})
\end{align*}
for some yet unknown functions $H, K, E, F, G,$ and $D$.
One can differentiate these equations and manipulate the equations to obtain
\begin{multline*}
 \left(HR^{-1}H-Q-\dot{H}\right){x}_j + \left(HR^{-1}K-\dot{K}\right)\lambda_j(t_i) \\+ \left(HR^{-1}E+Q\gamma_i^j-\dot{E}\right)=0
\end{multline*}
%%%%%%
and
\begin{multline*}
\left(FR^{-1}H-\dot{F}\right){x}_j + \left(FR^{-1}K-\dot{G}\right)\lambda_j(t_i)\\+\left(FR^{-1}E-\dot{D}\right)=0.
\end{multline*}
%%%%%%
In order to satisfy these equations for any value of ${x}_j(t)$ and $\lambda_j(t_i)$, the terminal conditions, and after noticing that $F = K^T$, we obtain that
\begin{align}
&\dot{H} = HR^{-1}H - Q, && H(t_i) = 0_{2\times2}\nonumber\\
&\dot{K} = HR^{-1}K, && K(t_i) = I_{2}\nonumber\\
&\dot{G} = K^TR^{-1}K, && G(t_i) = 0_{2\times2}\label{eq:Optimal_Tracking_Gains}\\
&\dot{E} = HR^{-1}E+Q\gamma_{i}^{j} , && E(t_i) = 0_{2\times1}\nonumber\\
&\dot{D} = K^TR^{-1}E , && D(t_i) = 0_{2\times1}.\nonumber
\end{align}
Note that $G(t_i) =0 $ and $\dot{G}(t)\succeq0$ for all $t$, which suggests that $G(t)\preceq0$ for $t<t_i$. If the problem is not \emph{abnormal}, i.e., there exists a neighboring minimum solution, then $G$ will be invertible at some $t<t_i$. In particular, by solving backwards in time in the sequence $H\to K\to E\to G\to D$ up to $t=t_{i-1}$, we can find that
\begin{multline*}
\lambda_j\left(t_i\right) =\\
%=
% G^{-1}\left(t\right)\left(\xi\left(i,j\right)-K^T\left(t\right){x}_j\left(t\right)-D\left(t\right)\right) \\
 G^{-1}\left(t_{i-1}\right)\left(\xi\left(i,j\right)-K^T\left(t_{i-1}\right){\xi}\left(i-1,j\right)-D\left(t_{i-1}\right)\right)
\end{multline*}
{resulting in the feedback optimal trajectory tracking control law}
\begin{multline}\label{eq:Braid_Optimal_Tracking_U_OL}
{u}_j^*(x_j,t)=
-{R}^{-1}\Big[H(t){x}_j(t)+K(t)G^{-1}(t_{i-1})\Big(\xi(i,j)\\
-K^T(t_{i-1})\xi(i-1,j)-D(t_{i-1})\Big)+E(t)\Big]
\end{multline}
where $H,K,E,G$ and $D$ are the solutions to terminal value problems in \eqref{eq:Optimal_Tracking_Gains}.
which can be solved numerically backwards from $t_i$. As it turns out, these conditions are also sufficient for optimality as presented in the following theorem.

%%%%%%%%%%%%%%%%%%%%%%%%%%%%%%%%%%%%%%%%%%%%%%%%%%%%%%%%%%%%%%%%%%%%%%%%%%%%%%%
\begin{figure*}[tbp]
\begin{center}
\subfloat[The robots start at the beginning of the braid.\label{fig:Rob_Start}]{\parbox{\textwidth}{\centering%
\scalebox{-1}[1]{\includegraphics[width=\linewidth,trim=40 210 0 160, clip]{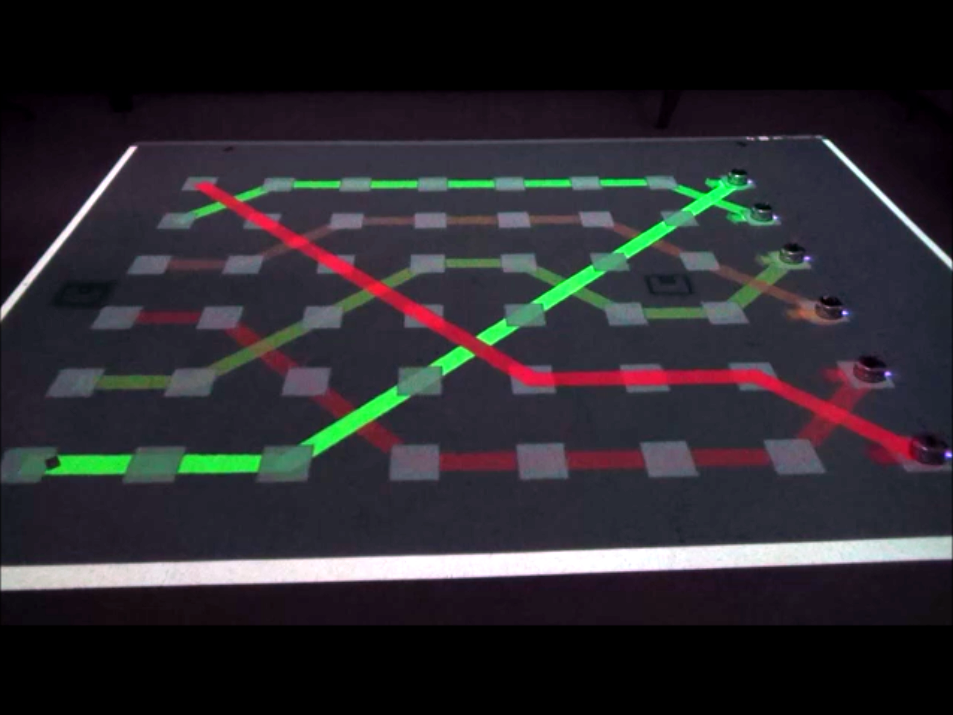}}%width=0.63212
}}\hfill
\subfloat[\emph{Collision-free} -- robots get as close as $\delta$.\label{fig:Rob_CollisionFree}]
{\parbox{\textwidth}{\centering%
\scalebox{-1}[1]{\includegraphics[width=0.495\textwidth,trim=40 210 0 160, clip]{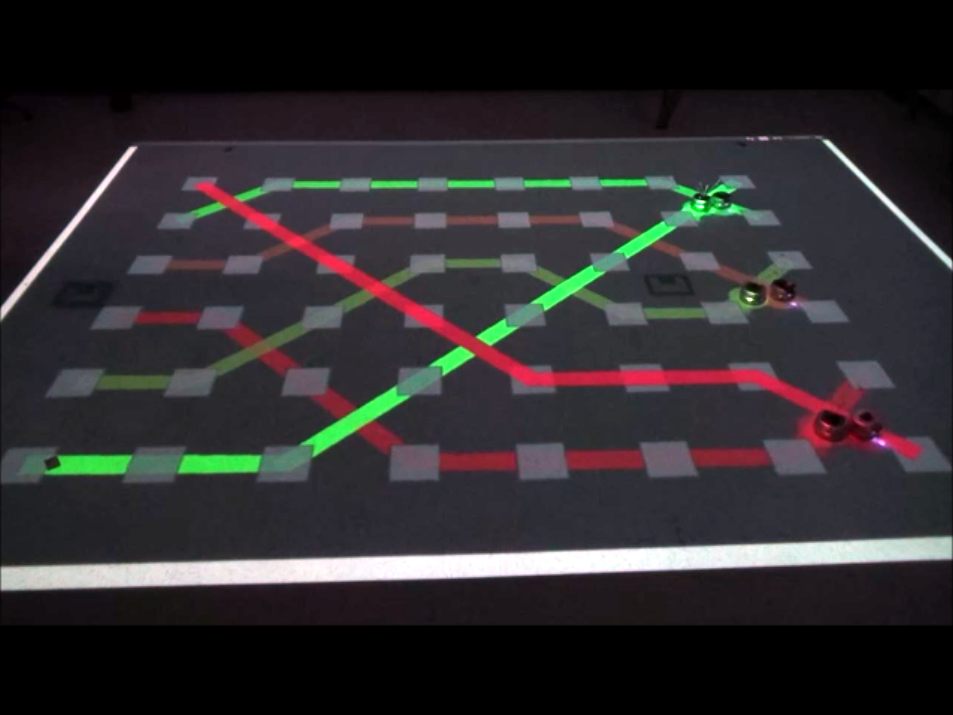}}\hfill
\scalebox{-1}[1]{\includegraphics[width=0.495\textwidth,trim=40 210 0 160, clip]{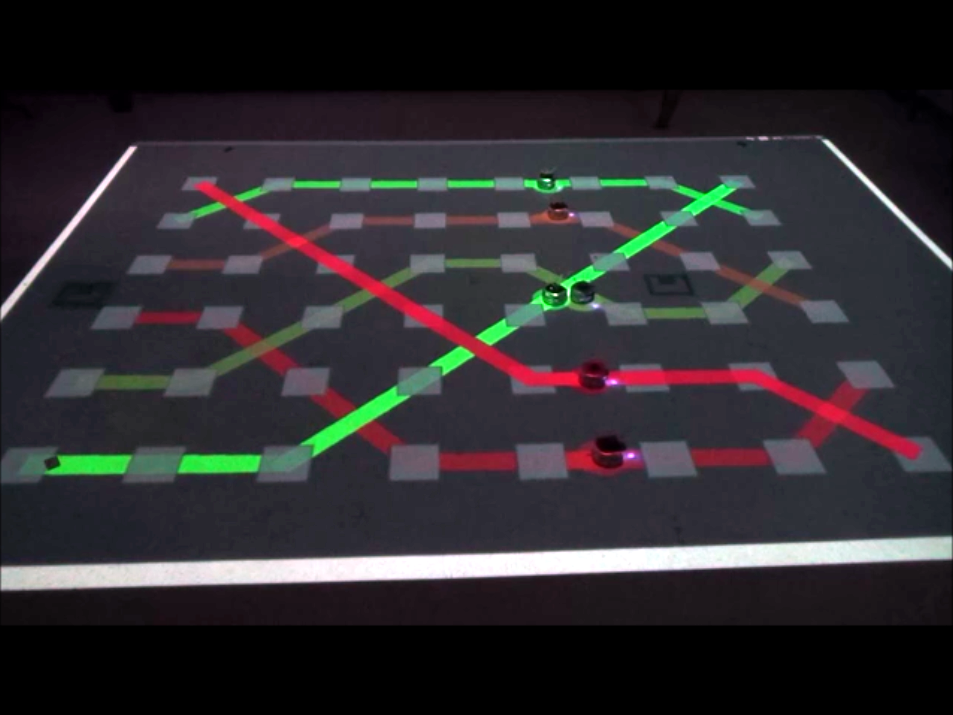}}
}}\\
\subfloat[The controller is \emph{braid point feasible} -- braid points are reached simultaneously.\label{fig:Rob_BraidPointFeasible}]
{\parbox{\textwidth}{\centering%
\scalebox{-1}[1]{\includegraphics[width=0.495\textwidth,trim=40 210 0 160, clip]{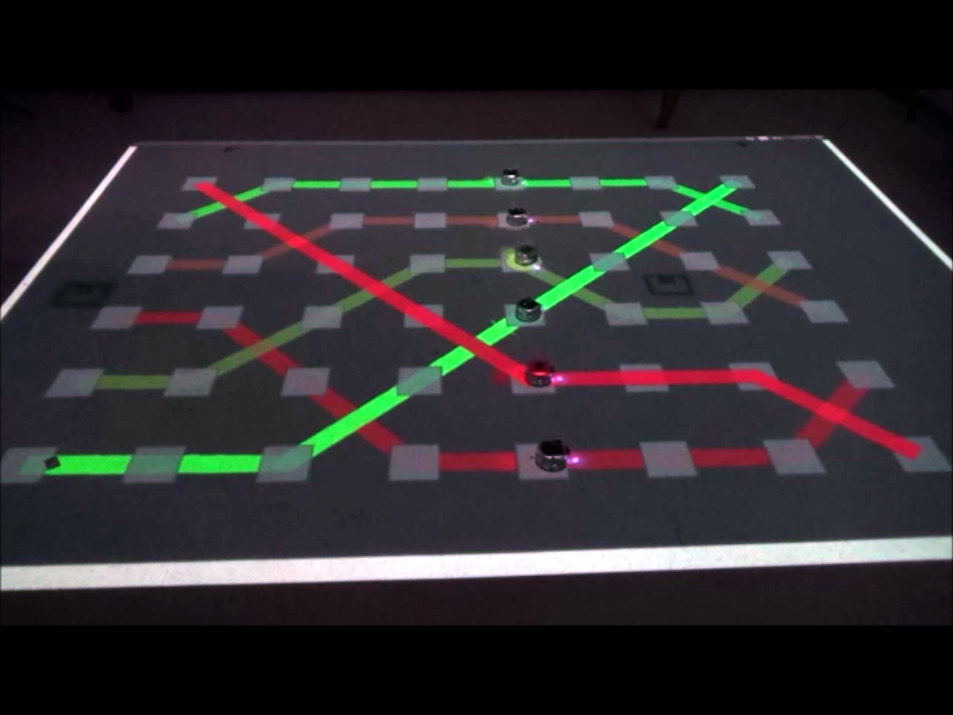}}\hfill
\scalebox{-1}[1]{\includegraphics[width=0.495\textwidth,trim=40 210 0 160, clip]{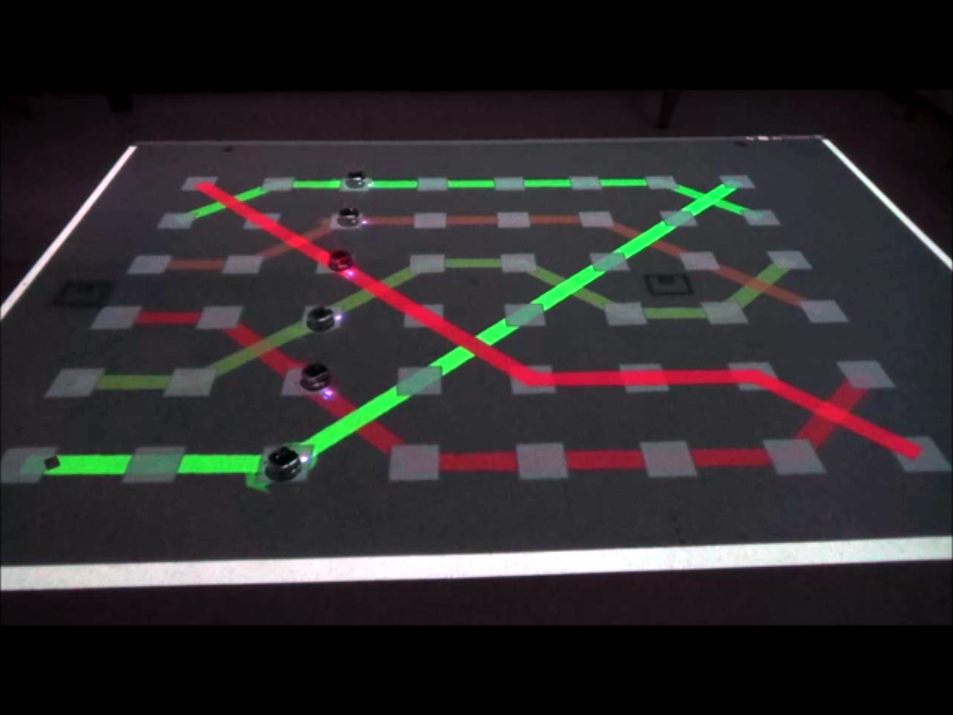}}
}}
\caption[Actual robots executing an inter-agent interaction strategy]{ Actual robots executing the mixing strategy in \eqref{eq:braid_string_implementation}. The geometric paths and spatio-temporal constraints are being projected on the workspace with an overhead projector for the sake of visualization.}
\label{fig:actualRobots}
\end{center}
\end{figure*}

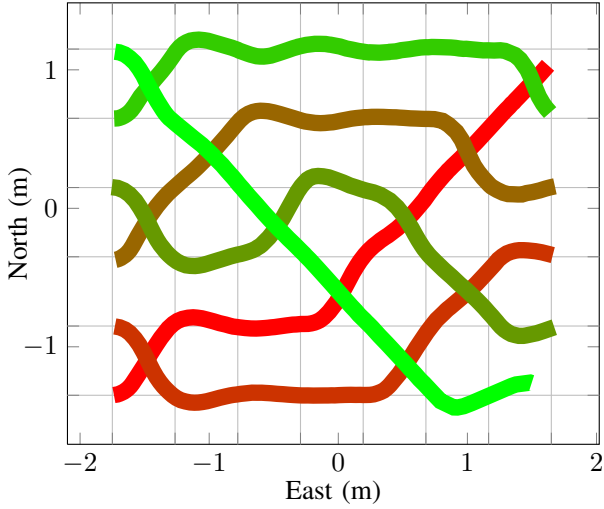
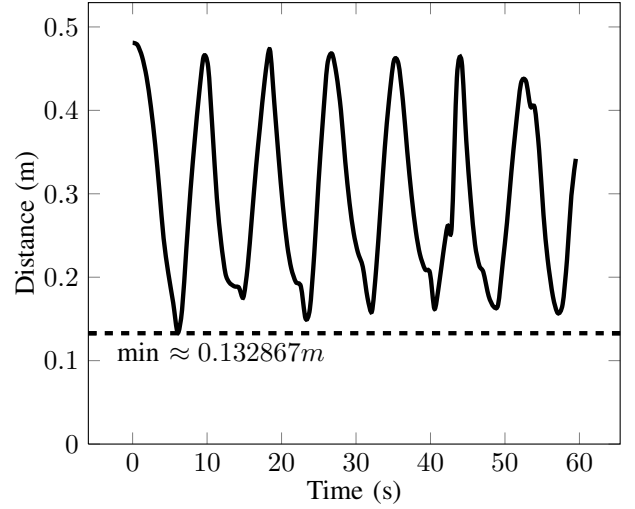
\begin{figure*}[t]
%\begin{center}
\subfloat[Robot trajectories in the plane.\label{fig:Rob_Trajectories}]%
{\parbox{0.99\columnwidth}{\centering%
\begin{tikzpicture}[scale=\linewidth/8.5cm]%
\begin{axis}[scale=1,xlabel=East (m), ylabel= {North (m)},%scale=0.55
			 compat=newest,
			 ylabel shift=-1em,%
			 xlabel shift=-0.45em,%
			 extra y ticks = {-1.35,-0.85,...,1.15},%
			 extra y tick labels = {,,,,,},
			 extra y tick style = {grid = major},%
			 extra x ticks = {-1.75,-1.2642857142857142857142857142857,...,1.6501},%
			 extra x tick labels = {,,,,,},%
			 extra x tick style = {grid = major}]%
\addplot+[no marks,red!100!green,line width=6pt,solid,%
		  x filter/.code=\pgfmathparse{-1*\pgfmathresult}]%
		 table[x=x1,y=y1,col sep=comma]{Compact_Braid_Data.csv};%
\addplot+[no marks,red!080!green,line width=6pt,solid,%
		  x filter/.code=\pgfmathparse{-1*\pgfmathresult}]%
		 table[x=x2,y=y2,col sep=comma]{Compact_Braid_Data.csv};%
\addplot+[no marks,red!060!green,line width=6pt,solid,%
		  x filter/.code=\pgfmathparse{-1*\pgfmathresult}]%
		 table[x=x3,y=y3,col sep=comma]{Compact_Braid_Data.csv};%
\addplot+[no marks,red!040!green,line width=6pt,solid,%
		  x filter/.code=\pgfmathparse{-1*\pgfmathresult}]%
		 table[x=x4,y=y4,col sep=comma]{Compact_Braid_Data.csv};%
\addplot+[no marks,red!020!green,line width=6pt,solid,%
		  x filter/.code=\pgfmathparse{-1*\pgfmathresult}]%
		 table[x=x5,y=y5,col sep=comma]{Compact_Braid_Data.csv};%
\addplot+[no marks,red!000!green,line width=6pt,solid,%
		  x filter/.code=\pgfmathparse{-1*\pgfmathresult}]%
		 table[x=x6,y=y6,col sep=comma]{Compact_Braid_Data.csv};%
\end{axis}%
\end{tikzpicture}}}%\hfill\ \!\!
\hfill%\quad \hfill%\\
\subfloat[Instantaneous minimum inter-robot distance.\label{fig:min_distance}]%
{\parbox{0.99\columnwidth}{\centering%\parbox{0.45\columnwidth}{
\begin{tikzpicture}[scale=\linewidth/8.5cm]%
\begin{axis}[scale=1,xlabel=Time (s),%scale=0.55
			 compat=newest,
			 ylabel style={align=center},%
			 ylabel=Distance (m),%
			 ylabel shift=-0.5em,%
			 xlabel shift=-0.35em,%
			 ymin = 0,%
			 ytick={0,0.1,0.2,0.3,0.4,0.5},%
			 extra y ticks={0.132867},%
			 extra y tick style={grid style={ultra thick,black},%
			 xtick align=outside,grid=major,dashed,%
			 tick style={ultra thick,black},%
			 tick label style={anchor=north west}},%
			 extra y tick labels={{~~min $\approx 0.132867 m$}}%
			 %tick style={white}
			 ]
\addplot[no marks, ultra thick, black,smooth] table[x=time,y=minimum_pairwise_distance,col sep=comma]{Compact_Braid_Data.csv};%
\end{axis}%
\end{tikzpicture}}}%
\caption[Braid controller robotic implementation data]{Data associated with robotic implementation in Fig. \ref{fig:actualRobots}.}
%\end{center}
\end{figure*}

\begin{theorem}
The tracking controller in \eqref{eq:Braid_Optimal_Tracking_U_OL} is a minimizer to the cost functional \eqref{eq:Tracking_Cost} whose optimal value is given by
\begin{multline*}
J(u^*) = 
\Big[\xi(i-1,j)^T\left(\frac{1}{2}H-KG^{-1}K^T\right)\xi(i-1,j)\\
+\xi(i-1,j)^T\left(KG^{-1}\left(\xi(i,j)-D\right)+E\right)+\varphi
\Big]\Bigg|_{t=t_{i-1}}
\end{multline*}
where $\varphi(t)$ is the solution to the terminal boundary problem
\begin{align*}
&\dot{\varphi} =\frac{1}{2}\left(\Lambda_j(t)\right)^TR^{-1}\Lambda_j(t) 
-\frac{1}{2}\left(\gamma_i^{j}(t)\right)^TQ\gamma_i^j(t)\\
&\varphi(t_i)  = -\xi(i,j)^T\Lambda_j(t_i)\xi(i,j)
\end{align*}
with
\begin{multline*}
\Lambda_j(t) = E(t)+\\
 K(t)G^{-1}(t_{i-1})(\xi(i,j)
-K^T(t_{i-1})\xi(i-1.j)-D(t_{i-1})).
\end{multline*}
\end{theorem}

\noindent\textit{Proof:}\newline\noindent
As the control law was derived from the necessary conditions for optimality, we only need to show that it is sufficient for optimality. We will do so by leveraging the Hamilton-Jacobi-Bellman theorem \cite[Chapter~2]{locatelli}.

Note that the optimization problem is regular as there exists a $u_j$ that allows the Hamiltonian to achieve a minimum with respect to it, i.e.,
\begin{multline*}
\mathcal{H}(x_j,u_j,\lambda_j)\\
=\frac{1}{2} \left[\left(x_j-\gamma_{i}^{j}\right)^TQ \left(x_j-\gamma_{i}^{j}\right)+u_j^TRu_j\right]+\lambda^T_ju_j\\
=\frac{1}{2} \left(u_j+R^{-1}\lambda_j\right)^TR\left(u_j+R^{-1}\lambda_j\right)-\frac{1}{2} \lambda^T_jR^{-1}\lambda_j\\
+\frac{1}{2}\left(x_j-\gamma_{i}^{j}\right)^TQ \left(x_j-\gamma_{i}^{j}\right)
\end{multline*}
which attains a minimum with respect to $u_j$ when $$u_j^*=-R^{-1}\lambda_j.$$
Define $V(z,t)$ as
\begin{multline*}
V(z,t) = \frac{1}{2}z^TH(t)z +z^TK(t)G^{-1}(t_{i-1})\xi(i,j)\\
-z^T\Big(K(t)G^{-1}(t_{i-1})\big(K^T(t_{i-1})\xi(i-1,j)-D(t_{i-1})\big)+E(t)\Big)\\
+\varphi(t)
\end{multline*}
where $\varphi(t)$ is as defined above.
It can be verified that $V(z,t)$ satisfies the terminal condition $V(\xi(i,j),t_i)=0$, that $\left.\frac{\partial V(z,t)}{\partial z}\right|_{z=x_j}^T=\lambda_j$, and that is satisfies the Hamilton-Jacobi-Bellman equation, i.e.,
\begin{multline*}
0=\frac{\partial V\left(z,t\right)}{\partial t}\Bigg|_{z=x_j}
\!\!\!\! \!\!\!\! \!\!\!\!
+\mathcal{H}\left(x_j,u_j^*\left(x_j,t\right),\frac{\partial V(z,t)}{\partial z}\Bigg|_{z=x_j}^T\right).
%,\frac{\partial V(z,t)}{\partial z}\Bigg|_{z=x_j}^T
\end{multline*}
As a consequence, $u_j^*$ is a minimizer to the cost functional and
\begin{align*}
J(u^*) = V(\xi(i-1,j),t_{i-1}).\tag*{$\blacksquare$}
\end{align*}

As a final note, 
the terminal costate value $\lambda_j(t_i)$ was computed using the initial conditions for the problem. However, as the gains involved are solved from terminal conditions, the choice of initial conditions is arbitrary, and evaluating at $t=t_{i-1}$ results in 
control law \eqref{eq:Braid_Optimal_Tracking_U_OL} being open-loop in the terminal costate value. This could yield undesired results under the influence of disturbances and errors. To alleviate this, we can rewrite the terminal costate as a function of the current state value instead, i.e.,
\begin{multline*}
\lambda_j(t_i) =\\
G^{-1}\left(t_{i-1}\right)\left(\xi\left(i,j\right)-K^T\left(t_{i-1}\right){\xi}\left(i-1,j\right)-D\left(t_{i-1}\right)\right) \\
= G^{-1}\left(t\right)\left(\xi\left(i,j\right)-K^T\left(t\right){x}_j\left(t\right)-D\left(t\right)\right)
\end{multline*}
in order to obtain the fully closed-loop optimal tracking controller
\begin{multline}\label{eq:Braid_Optimal_Tracking_U}
{u}_j^*(x_j,t)=-{R}^{-1}\left[\left({H}(t)-K(t)G^{-1}(t)K^T(t)\right){x}_j(t)\right.\\
\left.+K(t)G^{-1}(t)\left(\xi(i,j)-D(t)\right)+E(t)\right].
\end{multline}

%%%%%%%%%%%%%%%%%%%%%%%%%%%%%%%%%%%%%%%%%%%%%%%%%%%%%%%%%%%%%%%%%%%%%%%%%%%%%%%

\subsection{Robotic Implementation}

In order to validate the above results in a practical setting, the braid controllers were implemented on a team of Khepera III differential-drive robots, which may be modelled using unicycle dynamics, i.e.,
\begin{align*}
\dot{x}_j &= \begin{bmatrix}
\nu_j\cos\theta_j&\nu_j\sin\theta_j
\end{bmatrix}^T,&
\dot{\theta}_j &=\omega_j.
\end{align*}
where $x_j\in\R^2$ is the robot's position and $\theta_j$ its heading. The single integrator control $u_j$ from \eqref{eq:Braid_Optimal_Tracking_U} can be mapped to unicycle dynamics as
\begin{align*}
\nu_j &= \begin{bmatrix}
\cos\theta_j&\sin\theta_j
\end{bmatrix}\cdot u_j,\\
\omega_j &= \left\{\begin{array}{ll}
\kappa\left[\begin{smallmatrix}
-\sin\theta_j&\cos\theta_j
\end{smallmatrix}\right]\cdot\frac{{u}_j}{\norm{u_j}},&\mathrm{if}~\norm{u_j}>1
\\
\kappa\left[\begin{smallmatrix}
-\sin\theta_j&\cos\theta_j
\end{smallmatrix}\right]\cdot{u}_j,&\mathrm{otherwise}
\end{array}\right.
\end{align*}
for tuning gain $\kappa>0$.
% and where $\bar{u}_j = u_j\big/\|u_j\|$ if $\|u_j\|>1$ or $\bar{u}_j = u_j$ otherwise.
Fig. \ref{fig:actualRobots} illustrates the team of 6 real robots executing the braid string
\begin{align}\label{eq:braid_string_implementation}
\sigma = \left\{\sigma_1\!\cdot\!\sigma_3\!\cdot\!\sigma_5\right\}\!\cdot\!\sigma_2\!\cdot\!\sigma_3\!\cdot\!\sigma_4\!\cdot\!\left\{\sigma_3\!\cdot\!\sigma_5\right\}\!\cdot\!\left\{\sigma_2\!\cdot\!\sigma_4\right\}\!\cdot\!\sigma_1
\end{align}
where the braids grouped in braces 
are executed simultaneously.
A visual representation of the braid string geometry is being projected onto the robot workspace via an overhead projector mounted on the ceiling of the lab space. Fig. \ref{fig:min_distance} illustrates the minimum pairwise distance throughout the execution of the braid. It can be seen that the minimum distance achieved is approximately $0.132867 m$ -- greater than the size of the robots which is approximately $0.13m$. Fig. \ref{fig:Rob_Trajectories} illustrates the trajectories taken by the robots and how they follow the strand geometry, which were selected as straight lines, while still accommodating for the agents' unicycle dynamics.

%%%%%%%%%%%%%%%%%%%%%%%%%%%%%%%%%%%%%%%%%%%%%%%%%%%%%%%%%%%%%%%%%%%%%%%%%%%%%%%

\section{Conclusion}
The notion of multi-robot mixing is presented as a framework with which to characterize rich movement patterns in a multi-robot system symbolically in terms of inter-robot interactions. 
Two controllers are presented which allow to the multi-robot system to achieve provably safe interactions, in the sense that agents get close without colliding, while meeting certain spatio-temporal constraints. Under these controllers, it is possible to obtain theoretical bounds on the mixing limit, the greatest level of mixing or inter-robot interactions achievable in a given region. Further, tracking controllers that optimally follow desired geometries are provided which can be mapped to classes of nonholonomic dynamical systems. The concepts presented here are validated on an actual multi-robot system executing a interaction patterns with a desired mixing level.

Not found in this paper is the question of how to select the interaction pattern that should be executed by these controllers.
This question addresses the \emph{specification level} of our symbolic approach.
This discussion will be presented in a different paper where motion tasks are encoded in \emph{linear temporal logic}, allowing for syntactically rich specifications at a high level, such as ``agents 1 and 3 need to collected data before arriving at a goal location,'' which are then translated into a mixing strategy encoded as a braid.

\section*{Acknowledgment}

This work is partially supported under a grant from the
Air Force Office of Scientific Research.

The authors would like to thank the members of the Georgia Robotics and Intelligent Systems Laboratory (GRITS Lab) for the many fruitful discussions.

\ifCLASSOPTIONcaptionsoff
  \newpage
\fi

\IEEEtriggeratref{9} %<---------------------------breaks references

\bibliographystyle{IEEEtran}

\bibliography{lit_survey}%

\end{document}